\documentclass[9pt]{IEEEtran}
\usepackage{color}
\usepackage{docmute}
\usepackage{graphicx}
\usepackage{verbatim}
\usepackage{subfig}
\usepackage{MathDefs}
\usepackage{epstopdf}
\usepackage{csvsimple}
\usepackage{amsmath,amsfonts,amssymb,amsthm}
\usepackage{hyperref}
\hypersetup{
  colorlinks   = true, %Colours links instead of ugly boxes
  urlcolor     = blue, %Colour for external hyperlinks
 linkcolor    = blue, %Colour of internal links
  citecolor = blue,
}

\usepackage{tabularx}
\newcolumntype{L}[1]{>{\raggedright\let\newline\\\arraybackslash\hspace{0pt}}m{#1}}
\newcolumntype{C}[1]{>{\centering\let\newline\\\arraybackslash\hspace{0pt}}m{#1}}
\newcolumntype{R}[1]{>{\raggedleft\let\newline\\\arraybackslash\hspace{0pt}}m{#1}}

\usepackage{algorithm}
\usepackage{algpseudocode} % with end
\newenvironment{abbreviations}{\begin{list}{}{}}{\end{list}}

\begin{document}
\title{{A Critical Look at Coulomb Counting Towards Improving the Kalman Filter Based State of Charge Tracking Algorithms in Rechargeable Batteries}}

\author{
Kiarash Movassagh$^\dagger$,
Sheikh Arif Raihan$^\dagger$,
Balakumar Balasingam${^\star}{^\dagger},$ {\em Senior Member, IEEE}
and Krishna Pattipati$^\ddagger$, {\em Fellow, IEEE}
\thanks{Some preliminary findings were published in \cite{movassagh2019performance,movassagh2020performance}}
\thanks{Submitted to {\em IEEE Transactions on Control Systems Technology} in Jan. 2021. }
\thanks{$^\star$Balakumar Balasingam is the corresponding author at singam@uwindsor.ca}
\thanks{$^\dagger$Department of Electrical and Computer Engineering, University of Windsor, 401 Sunset Ave., Office\#3051, Windsor, ON N9B3P4, Canada, TP: +1(519) 253-3000 ext. 5431, E-mail: \{movassa,raihans,singam\}@uwindsor.ca.}
\thanks{$^\ddagger$Department of Electrical and Computer Engineering, University of Connecticut, 371 Fairfield Rd, Office\#350, Storrs, CT 06269, USA, TP: +1(860) 486-2890, E-mail: krishna.pattipati@uconn.edu.}
}

\maketitle

\begin{abstract}
In this paper, we consider the problem of state of charge estimation for rechargeable batteries. 
Coulomb counting is one of the traditional approaches to state of charge estimation and it is considered reliable as long as the battery capacity and initial state of charge are known.
However, the Coulomb counting method is susceptible to errors from several sources and the extent of these errors are not studied in the literature. 
In this paper, we formally derive and quantify the state of charge estimation error during Coulomb counting due to the following four types of error sources:
(i) current measurement error;
(ii) current integration approximation error; 
(iii) battery capacity uncertainty;
and
(iv) the timing oscillator error/drift. 
It is shown that the resulting state of charge error can either be of the {\em time-cumulative} or of {\em state-of-charge-proportional} type. 
Time-cumulative errors increase with time and has the potential to completely invalidate the state of charge estimation in the long run. 
State-of-charge-proportional errors increase with the accumulated state of charge and reach its worst value within one charge/discharge cycle. 
Simulation analyses are presented to demonstrate the extent of these errors under several realistic scenarios and the paper discusses approaches to reduce the time-cumulative and state of charge-proportional errors. 

\end{abstract}


\section{Introduction}
\label{sec:intro}
Rechargeable batteries are becoming an integral part of the future energy strategy of the globe. 
The use of rechargeable batteries are steadily on the rise in a wide ranging applications, such as,
electric and hybrid electric vehicles, household appliances, robotics, power equipment, consumer electronics, aerospace, and renewable energy storage systems.
Accurate estimation of the state of charge (SOC) of a battery is critical for the safe, efficient and reliable management of batteries \cite{balasingam2018elements,waag2013line,plett2004extended,yuan2013state}.

There are {three} approaches to estimate the SOC of a battery \cite{balasingam2020BMSenergies}:
(i) voltage-based approach,
(ii) current-based approach,
{and
(iii) fusion of voltage/current based approaches. 
The fusion-based approaches seek to retain the benefits of both voltage-based and current-based approaches by employing non-linear filters, such as the extended Kalman filter, in order to fuse the information obtained through the voltage and current measurements. 
}

In its simplest form, the voltage-based approach serves as a table look-up method --- the measured voltage across the battery terminals is matched to its corresponding SOC in the OCV-SOC characterization curve  \cite{pattipati2014open}. 
{
In more generic terms, we have 
\begin{align}
\text{voltage-measurement} = 
\underbrace{f({\rm SOC})}_{\text{OCV-SOC model}} + 
\underbrace{g(\text{parameters, current})}_{\text{voltage drop}}
\label{eq:MEAS-generic}
\end{align}
where the function $f(\cdot)$ refers to the {\em open circuit voltage model} that relates the OCV of the battery to SOC and 
 $g(\cdot)$ accounts for the voltage-drop within the battery-cell due to hysteresis and relaxation effects.   
Challenges in voltage-based SOC estimation arise due to the fact 
that the functions $f(\cdot)$ and $g(\cdot)$ are usually non-linear and that there is a great amount of uncertainty as to what those functions might be  \cite{pattipati2014open,balasingam2014robust}. 
For instance, the parameters can be modelled through electrical equivalent circuit models (ECM) \cite{plettBMSbookpart1,plettBMSbookpart2} or electrochemical models \cite{hariharan2017mathematical} each of which can result into numerous reduced-order approximations. 
The voltage based approach suffers from the following three types of errors:
\begin{enumerate}
\item[(i)] {\em OCV-SOC modeling error.}
The OCV-SOC relationship of a battery can be approximated through various models \cite{pattipati2014open}: 
linear model, polynomial model and combine models are few examples. 
Reducing the OCV-SOC modeling error is an ongoing research problem --- in  \cite{ahmed2020scaling} a new modeling approach was reported that resulted in the ``worst case modeling error'' of about 10 mV.
It must be mentioned that the OCV modeling error is not identical at all voltage regions of the battery. 
\item[(ii)] {\em Voltage-drop modeling error.}
Voltage-drop models accounts for the hysteresis and relaxation effects in the battery. 
Various approximations were proposed in the literature in order to represent these effects \cite{nikdel2014various,balasingam2014robust}. 
\item[(iii)] {\em Voltage measurement error.}
Every voltage measurement system comes with errors; this translates into SOC estimation error. 
\end{enumerate}
In order to reduce the effect of uncertainties in voltage-based SOC estimation, it is often suggested to rest the battery before taking the voltage measurement for SOC lookup \cite{plettBMSbookpart2} --- when the current is zero for sufficient time the voltage-drop also approaches to zero.
However, all the other sources of errors mentioned above (OCV-SOC modeling error, hysteresis, and voltage measurement error) cannot be eliminated by resting the battery. 
}

The current-based approach, also known as the {\em Coulomb counting method}  \cite{plettBMSbookpart2}, computes the amount of Coulombs added/removed from the battery in order to compute the SOC as a ratio between the remaining Coulombs and the battery capacity that is assumed known. 
{The Coulomb counting approach to SOC estimation can be approximated as follows (see Section \ref{sec:probDef} for details)
\begin{align}
{\rm SOC}(k) &= {\rm SOC}(k-1) + \frac{\Delta_k  i(k) }{3600 {\rm C}_{\rm batt}} 
\label{eq:CC-generic}
\end{align}
where 
${\rm SOC}(k)$ indicates the SOC at time $k$,
$\Delta_k$ is the sampling time, 
$i(k)$ is the current through the battery at the at time $k$, and 
${\rm C_{batt}}$ is the battery capacity an Ampere hours (Ah).
}
Assuming the knowledge of the initial SOC, the Coulomb counting method computes the effective change in Coulombs in/out of the battery based on the measured current and time in order to compute the updated SOC. The important advantage of the Coulomb counting approach is that it does not require any prior characterization, such as the OCV-SOC characterization \cite{pattipati2014open} that is required for the voltage-based SOC estimation method. 
However, the Coulomb counting method can result in SOC estimation errors due to the following five factors:
\begin{enumerate}
\item {\em Initial SOC.}
Coulomb counting approach assumes the knowledge of the initial SOC before it starts counting Coulombs in and out of the battery based on the measured current.  
Any error/uncertainty in the initial SOC will bias the Coulomb counting process.  
\item {\em Current measurement error.}
Current sensors are corrupted by measurement noise \cite{wang_combined_2007, ng_enhanced_2009}; simple, inexpensive current sensors are likely to be more noisy. 
\item {\em Current integration error.}
Coulomb counting methods employ a simple, rectangular approximation for current integration. 
Such an approximation results in errors that increase with sampling as the load changes rapidly. 
\item {\em Uncertainty in the knowledge of battery capacity \cite{balasingam2014robustP2}.}
Coulomb counting method assumes perfect knowledge of the battery capacity, which is known to vary with temperature, usage patterns and time (age of the battery) \cite{balasingam2015performance,avvari2015experimental}. 
\item {\em Timing oscillator error.}
Timing oscillator provides the clock for (recursive) SOC update, i.e., the {\em measure of time} comes from the timing oscillator.
Any error/drift in the timing oscillator will have an effect on the measured Coulombs. 
\end{enumerate}

{
The fusion-based approach seeks to retain the best features of both voltage and current-based approaches. 
This is achicved by creating the following {\em state-space model} 
\begin{align}
{\rm SOC}(k) &= {\rm SOC}(k-1) + \frac{\Delta_k  i(k) }{3600 {\rm C}_{\rm batt}} + n_{\rm s} (k) \label{eq:process-eq} \\
z_{\rm v} (k) &=
\underbrace{ f({\rm SOC}(k))}_{\text{OCV-SOC model}}+ 
\underbrace{g(\text{parameters}, i(k))}_{\text{voltage drop}} + n_{\rm z}(k) \label{eq:meas-eq}
\end{align}
where 
\eqref{eq:process-eq} is the {\em process model} that is derived from the Coulomb counting equation \eqref{eq:CC-generic},
\eqref{eq:meas-eq} is the {\em measurement model} that is is derived from the voltage measurement equation \eqref{eq:MEAS-generic}, 
$n_{\rm s} (k)$ is the process noise, 
$z_{\rm v} (k)$ is the measured voltage across the battery terminals, 
$ f ({\rm SOC}(k))$ is the OCV characterization function \cite{pattipati2014open} that represents the battery voltage as a function of SOC, 
$g(\text{parameters}, i(k))$ is the voltage drop due to impedance and hysteresis within the battery, 
and
$ n_{\rm z}(k)$ is the measurement noise corresponding to the model \eqref{eq:meas-eq}. 
The goal from the above state-space model is to recursively estimate the SOC given the voltage and current measurements. 
}

{
The state-space model described in \eqref{eq:process-eq}-\eqref{eq:meas-eq} is non-linear due to the OCV-SOC model and the different approximate representations for voltage-drop models \cite{balasingam2014robust}. 
If the models are known, a non-linear filter, such as the extend Kalman filter \cite{bar2004estimation}, can yield near-accurate estimate of SOC in real time.
The filter selection is based on the model assumptions:
\begin{itemize}
\item[(I)] {\em Kalman filter:-}
Here, the following assumptions need to be met:
the state-space model is known and linear, i.e., the functions  $f(\cdot)$ and $g(\cdot)$ in \eqref{eq:meas-eq} are linear in terms of SOC, 
the 'parameter' and the current $i(k)$ in \eqref{eq:meas-eq} are known with negligible uncertainty in them. 
and that the process and measurement noises, $n_{\rm s}(k)$ and $n_{\rm v}(k)$, respectively, are i.i.d. Gaussian with known mean and variance. 
\item[(II)] {\em Extended/unscented Kalman filter:-}
Here, only the linearity assumption is relaxed, i.e., the functions  $f(\cdot)$ and $g(\cdot)$ in \eqref{eq:meas-eq} can be non-linear in terms of SOC. 
All other assumptions for the Kalman filter need to be met, i.e., the model parameters and the noise statistics need to be perfectly known and that the  process and measurement noises need to be i.i.d. Gaussian with known mean and variance. 
\item[(III)] {\em Particle filter:-}
Compared to the Kalman filter assumptions, particle filter allows to relax both linear and Gaussian assumptions; here, the $f(\cdot)$ and $g(\cdot)$ can be non-linear and both process and measurement noise statistics can be non-Gaussian. 
It needs to be re-emphasized that, similar to the cases in (I) and (II) above, the models  $f(\cdot)$ and $g(\cdot)$ and the parameters of the noise statistics need to be perfectly known. 
\end{itemize}
}

{
It is important to note that the recursive filters discussed above all assume that the model, which consists of the functions $f(\cdot)$, $g(\cdot)$ and the parameters of the noise statistics, is perfectly known. 
However, we have discussed several ways earlier in this section in which the known-model assumptions can be violated. 
Indeed, the ``known model'' assumptions can be violated through any of the following ten ways:
\begin{itemize}
\item[(a)] {\em Five sources of error in defining the process model \eqref{eq:process-eq}:}
namely, the
initial SOC error, 
current measurement error, 
current integration error,
battery capacity error,
and timing oscillator error. 
\item[(b)] {\em Three sources of error in defining the measurement model \eqref{eq:meas-eq}:}
namely,
OCV-SOC modeling error,
voltage-drop modeling and its parameter estimation error, 
and
voltage measurement error.
\item[(c)] {\em The process noise $n_{\rm s} (k)$.}
The statistical parameters of the process noise should be computed based on the knowledge about the statistics of the five error sources in (a) above. 
\item[(d)] {\em The measurement noise $n_{\rm v} (k)$.}
The statistical parameters of the measurement noise should be computed based on the knowledge about the statistics of the three error sources in (b) above. 
\end{itemize}
The focus of the present paper is to develop detailed insights about the error sources (a) and (c) above.
In a separate work \cite{ECM0_2020}, we discuss the noise sources (b) and (d) in detail. 
}

\subsection{Background}
\label{sec:literature}

{
The classical estimation theory \cite{bar2004estimation} states that when the linear-Gaussian conditions and the known model assumptions (stated under (I) in Section \ref{sec:intro}) are met the SOC estimate will be {\em efficient}, i.e., the variance of the SOC estimation error will be equal to that of the \textit{posterior Cram\'er-Rao lower bound (PCRLB}) which is proved to be the theoretical bound; under non-Bayesian conditions this limit is known just as the Cram\'er-Rao lower bound (CRLB).
That is, the PCRLB or CRLB can be used as a {\em gold standard} on performance.}
In this regards, some prior works in the literature have \cite{lee2019estimation,lin2018theoretical,bizeray2018identifiability} derived the CRLB as a measure of performance evaluation.  
These approaches were developed to estimate the ECM parameters of the battery; the ECM parameters are involved in the measurement model for the SOC in \eqref{eq:MEAS-generic}. 
In addition to the use in SOC estimation, ECM parameter estimation has other important applications in a battery management system.

Several other approaches attempted to theoretically derive the error bound on SOC estimation separately and jointly with ECM parameter identification. 
In \cite{song2018parameter}, a RLS-based parameter identification technique with forgetting factor was presented in which a sinusoidal current excitation made of two sinusoid component was used.  According to the results, the CRLB of resistance decrease with the increase of frequencies and thus the large frequency components are preferable for higher accuracy in parameter estimation; similar observations were reported in \cite{song2019combined}. 
Influence of voltage noise, current amplitude and frequency on parameter identification has been illustrated in \cite{song2018parameterCRLB} where a sinusoidal excitation current was used. 
Here, the CRLB of the battery equivalent circuit model was derived using Laplace transform; the authors round no influence of the frequency of the excitation signal on the single-parameter identification of ohmic resistance and reported that reducing the voltage measurement noise and increasing current amplitude improves the identification accuracy.
A posterior CRLB was developed to quantify accuracy for EKF based ECM parameter identification in which a second order battery ECM was adopted in \cite{klintberg2017theoretical}. 
The CRLB was determined numerically with the help of sinusoidal current excitation. 
It showed that the CRLB of ohmic resistance estimation decreases with the increase of current amplitude and frequency as well.
Unlike \cite{klintberg2017theoretical}, the CRLB was derived in analytic expression in discrete time and Laplace transform in \cite{lin2017analytic} in which a (known) sinusoidal current input was considered. 
A non-linear least-square based electrode parameter (e.g. electrode capacity) identification method was presented in \cite{lee2019estimation} in which only the terminal voltage was considered to contain measurement noise. 
This CRLB was derived and used to quantify the error bound of the estimator to determine the uncertainty of the parameter estimation. 
The parameter estimates were interpreted with the help of analytically derived \textit{confidence levels}. 
Here, the noise was assumed to be Gaussian white noise with standard deviation 10 mV in the demonstrations.
In \cite{lin2018theoretical}, battery SOC estimation error was derived theoretically as a function of sensor noises; the proposed approach considers measurement noise in both current and voltage. 
Effect of different components involved in SOC estimation were demonstrated using a parameter sensitivity analysis in \cite{zhang2014parameter}
and the effect of bias and noise were reported in \cite{mendoza2017relative} as well.

The five sources of error in Coulomb counting have been recognized in the literature and some remedies were proposed. 
In \cite{Zhang2014}, the initial SOC is modeled as a function of the terminal voltage, temperature and the relaxation time.
The authors in \cite{Yang} proposed the use of neural networks to gain a better estimate of the initial SOC.
In \cite{Yan2010} a data fusion approach is proposed where a H-infinity filter is used to minimize the error in the initial SOC estimate.
In the battery fuel gauge evaluation approach proposed in \cite{avvari2015experimental} the uncertainty in initial SOC error was taken into account and the OCV lookup method  \cite{pattipati2014open,xiong2017novel} is introduced as a performance metric. 
It was pointed out in \cite{dong2016online} that the accuracy of the OCV lookup method might be affected with battery age. 
The effect of current integration error was also recognized in the literature and remedies were proposed:
in \cite{Cho,Yang} a model based approach was proposed to reduce current integration error;
in  \cite{Wu2017}, it was proposed to reset Coulomb counting when the present SOC is known when the battery is fully charged/discharged where the ``fully-charged'' and ``fully-empty'' conditions were declared based on measured voltage across battery terminals; here, the authors propose a way to minimize the error due to voltage-only based declaration of these two conditions.
Many articles recognize the imperfect knowledge of battery capacity and ways to estimate them;
a neural network based approach to battery capacity estimation was proposed in in \cite{Yang}; 
an approach based on the charge/discharge currents and the estimated SOC for battery capacity estimation was proposed \cite{SEPASI2015}; 
the authors in \cite{Wu2017} propose estimating the battery capacity when the battery is fully charged/discharged, which can be known easily when the terminal voltage reach the max./min. voltage respectively;
in \cite{balasingam2014robustP2} a state-space model was introduced to track battery capacity where measurements can be incorporated using multiple means, including when the battery is at rest.  
None of the existing works explored the effect of timing oscillator error in the estimated SOC. 

In summary, the importance of theoretical performance derivation and analysis is recognized in the literature, particularly in the above detailed publications. 
Considering the nature of the complexity of the real-world measurement model, the existing literature represents only a small fraction of what needs to be done for a complete understanding of the battery SOC estimation problem. 
{
For example, even though the effect of some of the five sources of Coulomb counting error (summarized earlier in this section) were noted in the literature, it was not fully incorporated into the fusion-based SOC tracking approaches. 
In other word, the process noise $ n_{\rm s} (k)$ in \eqref{eq:process-eq} was not accurately defined in the literature. 
Table \ref{table:process-noise} summarizes how the process noise is defined in some notable works in the literature. 
Setting arbitrary values to a process noise will have the following adverse effect on the filter outcome:
\begin{itemize}
\item {\em Too small process noise:}
When the process noise is smaller than the reality, the filter will compute the weights such that the measurements are ignored. 
\item {\em Too large process noise:}
When the process noise is larger than the reality, the variance of the filtered estimates will be high -- effectively the benefits of using a filter will be lost. 
\end{itemize}
Based on Table \ref{table:process-noise}, it is clear that there is a knowledge gap about the process noise in recursive-filtering approach to SOC tracking. 
The focus of this paper is to derive accurate models for SOC tracking; particularly we focus on the process model only. 
Similar discussion about one of the possible measurement models can be found in \cite{ECM0_2020}. 
Model validation strategies and analyses using practical data are left for a future discussion. 
\begin{table*}[h]
\begin{center}
\caption{{\bf Process noise in SOC tracking}}
\label{table:process-noise}
{
\begin{tabular}{|C{.8 in}|C{1.2 in}|C{4 in}|}
\hline
{\bf Paper} & {\bf Filtering Method} & {\bf Definition of process noise} \\
\hline
\cite[page 279]{plett2004extended-p3}& Extended Kalman filter & ``small'' \\
\hline
\cite[page 1370]{plett2006sigma} & Unscented Kalman filter  & ``stochastic process noise or disturbance  that models some unmeasured input which affects  the  state  of  the  system''\\
\hline
\cite[page 7]{linghu2019} & Kalman filter & ``process noise'' \\
\hline
\cite[page 334]{wei2016} & Frisch scheme based bias  & ``zero-mean white noise with variance $\sigma_i$''\\
& compensating recursive least squares &  \\
\hline
\cite[page 8954]{wadi2019} & Extended Kalman filter & ``zero-mean white Gaussian process noise''  \\
\hline
\cite[page 13205 \& 13206]{peng2017} & Adaptive unscented Kalman filter & ``zero-mean Gaussian white sequence''; ``In practice, the mean and covariance of process noise is frequently unknown or incorrect'' \\
\hline
\cite[page 4610]{el2015enhancement} & Extended Kalman filter & ``The EKF assumes knowledge of the measurement noise  statistics. Moreover,\\
 & & any uncertainty in the system’s model will degrade the estimator’s performance'' \\
 \hline
\cite[page 10]{sun2018adaptive}& Correntropy unscented Kalman filter &  ``The process noise covariance and measurement noise are assumed to be known in CUKF. However, they are real time in general and may not be obtained prior in practice. Therefore, they should be updated with changes in time on the basis of some obtained prior knowledge.''\\
\hline
& & ``$ w_k \sim \mathcal{N}( 0 ,\,\bQ_k)$'' where $\bQ_k$ is the covariance matrix\\
\cite[page 166660]{zhang2019state} & Adaptive weighting Cubature particle filter & ``In the process of practical application, the statistical characteristics of the process noise and measurement noise of the system are highly random and vulnerable to external environmental factors.'' \\
\hline
\cite[page 8614]{xi2019learning}& Extended Kalman filter & ``Model bias is the inherent inadequacy of the model for representing the real physical systems due to the model assumptions and simplifications.''\\
\hline
\cite[page 5,8]{bi2020adaptive} & Adaptive square-root sigma-point Kalman filter & ``$w_k$ refers to process noise, which represents unknown disturbances that affect the state of the system''; ``Usually, covariance matrices are constant parameters determined
offline before the estimation process begins. In practice, the characteristics of noises vary depending on the choice of sensors and the operating conditions.''\\
\hline
\end{tabular}}
\end{center}
\end{table*}
}

\subsection{Summary of Contributions}

A large portion of the existing work related to battery SOC estimation in the literature lack theoretical validations. 
Almost all the work that employ some form of theoretical validation are summarized in Subsection \ref{sec:literature} --- the number of papers in this section is insignificant compared to the number of publication in SOC estimation in the past year alone. 
This indicates the need to focus more in theoretical performance analyses and to understand where the remaining challenges in battery SOC estimation. 
 
In this paper, we develop a mathematical model to theoretically compute the accumulated SOC error as a result of 
current measurement error,
current integration approximation,
battery capacity uncertainty, and 
timing oscillator error. 
These four sources of error are identified in \cite{KiaEPEC}. In this paper, we provided the formulas for exact statistical error parameters (mean and standard deviation) that can be used to improve all existing SOC estimation methods. 
As such, the contributions of this paper are summarized as follows:
\begin{itemize}
\item 
{\em Exact computation of Coulomb counting error.}
With realistic numerical examples, we demonstrate the errors and their severity during Coulomb counting. 
Further, we derive mathematical formulas to determine these errors such that the statistical confidence in the SOC estimates can be explicitly stated. 

\item 
{\em {Five} different error sources in Coulomb counting are analyzed.}
We derive the exact mean and standard deviation of the error (with time) due to all {five} possible sources of errors during Coulomb counting:
current measurement error, 
current integration error, 
battery capacity uncertainty,
{charge,} {discharge efficiency uncertainty},
and timing oscillator error.
It is demonstrated that the resulting error will fall into one of the following two categories:
{time-proportional errors} and {\em SOC-proportional errors.}

\item 
{\em Time proportional errors increase indefinitely.}
We demonstrate that the standard deviation of the time-proportional error approaches to infinity as the number of samples reaches to infinity. 

\item 
{\em State of charge proportional errors reach worst case within one cycle.}
It is shown that the errors due to battery capacity uncertainty and timing oscillator drifts reach their peak values within one discharge/charge cycle.
In addition, the standard deviation of these errors vary with the accumulated SOC.
The proposed exact model can be used to improve the SOC estimation by incorporating them in state space models, e.g., the proposed model can be used to improve the extended Kalman filter based SOC estimation techniques \cite{plett2004extended}.

\item 
{\em Accurate state-space models for real-time state of charge estimation.}
The models were presented in a way that their applicability in state-space models is explicit. 
The proposed models can be used to improve the accuracy of virtually all online filtering approaches, i.e.,
those based on extended Kalman filter, unscented Kalman filter, particle filter etc., that have been employed for real-time SOC estimation.

\end{itemize}
The effect of initial SOC error will remain as a bias in the Coulomb counting process, and as such it does not require any further analysis in this paper. 
Some initial versions of the derivations presented in this paper were reported in \cite{movassagh2019performance}; the present papers expands all derivations presented \cite{movassagh2019performance} towards a generalized state-space model.

{
It must be noted that all the contributions listed above will translate into an accurate process noise model in the state-space model for recursive SOC tracking.
It will be shown later in this paper that the process noise variance is a significantly time-varying quantity --- something never considered in the literature before.  
Further, even though Coulomb counting is considered an outdated approach to SOC estimation, it is still widely used in practical implementations \cite{balasingam2020batterypatent,balasingam2015performance,avvari2015experimental}.
For example, whenever the fusion based approaches encounter failures, due to unexpected measurements and errors etc., the battery management systems are usually programmed to fall back to the Coulomb counting method as an alternative. 
Hence, the paper is written in a way that quantifies the error in computed SOC from Coulomb counting. 
Later, we discuss how the findings in this paper will be used to derive an accurate model for voltage-current fusion based SOC tracking using recursive filters. 
}

\subsection{Organization of the Paper}

The remainder of this paper is organized as follows:
Section \ref{sec:probDef} formally introduces Coulomb counting and identifies the four different error sources. 
The accumulated error in SOC due to 
current measurement error,
current integration approximation,
battery capacity uncertainty, 
and timing oscillator drift are derived and analyzed in 
Sections \ref{sec:CMerror}, \ref{sec:CIerror},  \ref{sec:CbattError} and  \ref{sec:TimingError}, respectively. 
{A summary of individual uncertainties and their effect on the counted Coulombs is presented in in Section  \ref{sec:summary}. 
In Section \ref{sec:combined}, some practical ways are discussed into how individual effects can be combined into the process model of a recursive filter implementation for SOC tracking. 
Finally, the paper is concluded in Section \ref{sec:concl}.}

\section*{List of Acronyms} 
\begin{abbreviations}
\item[CRLB] Cramer-Rao lower bound
\item[ECM] Equivalent circuit model
\item[EKF] Extended Kalman filter
\item[OCV] Open circuit voltage 
\item[PCRLB] Posterior Cramer-Rao lower bound
\item[RLS] Recursive least squares
\item[SOC] State of charge
\end{abbreviations}

\section*{List of Notations} 
List of notations used in the remainder of this paper are summarized below.
\begin{abbreviations}

\item[$\rm C_{true}$] True battery capacity (see \eqref{eq:cbatt-ctrue})
\item [${\rm C}_{\rm batt}$] Assumed battery capacity \eqref{eq:CCct}
\item[$\rm C_{\Delta}$] Battery capacity uncertainty \eqref{eq:cbatt-ctrue}

\item[$ \delta_{\rm I}(k)$] Current integration error at time $k$ \eqref{eq:int_error}
\item [$\Delta_k $] Sampling duration at time $k$ \eqref{eq:ccapprox0}
\item [$ \Delta$] Sampling time that is assumed constant \eqref{eq:Delta}
{
\item[$\Delta_{\rm true}$] True sampling time \eqref{time:error}
\item[$\Delta_{\epsilon}$] Timing oscillator error \eqref{time:error}
\item [$ \eta$] Coulomb counting efficiency \eqref{eq:CCct} 
\item [$ \eta_\rc$] Charging efficiency \eqref{eq:ch/dc-efficiency} 
\item [$ \eta_d$] Discharging efficiency \eqref{eq:ch/dc-efficiency} 
}

\item [$i(t)$] Current through battery at time $t$ \eqref{eq:CCct}
\item [$i(k)$] Sampled current through battery at time instant $k$ \eqref{eq:CCcontin}

\item [$n_{\rm i}(k)$] Current measurement noise \eqref{eq:z_i}
\item [$n_{\rm s}(k)$] Process noise \eqref{eq:process-eq}
\item [$n_{\rm z}(k)$] Measurement noise \eqref{eq:meas-eq}

\item[$\kappa$] Integration error constant \eqref{eq:int_err_sigI}

\item[${\rm \rho_i} $] Current measurement noise coefficient \eqref{eq:rho_i}
\item[${\rm \rho_I} $] Current integration noise coefficient \eqref{eq:rho_I}
\item[${\rm \rho_C} $] Capacity uncertainty coefficient \eqref{eq:rhoC}
{
\item[$\rho_{\eta_\rc}$] Charging uncertainty coefficient \eqref{eq:rho-charging-efficiency}
\item[$\rho_{\eta_\rd}$] Discharging uncertainty coefficient \eqref{eq:rho-charging-efficiency}
\item[$\rho_\Delta $] Timing error coefficient \eqref{eq:TEcoefficient}
}

\item [$s(t)$] SOC at time $t$ \eqref{eq:CCct}
\item [$s(0)$] Initial SOC \eqref{eq:CCct}
\item [$s(k)$] SOC at discretized time instance $k$ \eqref{eq:CCcontin}
\item[$s_{\rm CC}(n)$] Change in SOC over $n$ samples \eqref{eq:CCerrorSOC_n}
\item[$ \sigma_{\rm i}$] Std. deviation  of current measurement error \eqref{eq:sig_i}
\item[$\sigma_L$] Std. deviation  of load current changes \eqref{eq:int_propto_sigL}
\item[$\sigma_{\rm batt}$] Std. deviation of battery capacity uncertainty \eqref{eq:CbattError}
{
\item[$\sigma_{\eta_\rc}$] Std. deviation of charging uncertainty \eqref{eq:STD-c/d-coeff}
\item[$\sigma_{\eta_\rd}$] Std. deviation of discharging uncertainty \eqref{eq:STD-c/d-coeff}
\item[$\sigma_{\Delta}$] Std. deviation of timing uncertainty \eqref{eq:soc-err-combined-var}
}
\item[$\sigma_{\rm s, i} (n)$] Std. deviation of $w_{\rm i}(n)$ \eqref{eq:STD-imeas-error}
\item[$\sigma_{\rm s, I} (n)$] Std. deviation  of $w_{\rm I}(n)$ \eqref{eq:STD-integ-error}
\item[$\sigma_{\rm s, C} (n)$] Std. deviation of $w_{\rm C}(n)$ \eqref{eq:STD-Cbatt-error}
{
\item[$\sigma_{\rm s, \eta} (n)$] Std. deviation of $w_\eta(n)$ \eqref{eq:STD-c/d-coeff}
\item[$\sigma_{\rm s, \Delta} (n)$] Std. deviation of $w_\Delta (n)$ \eqref{eq:err-var-del}
\item[$\sigma_{\rm s} (n)$] Std. deviation of $w (n)$ \eqref{eq:soc-err-combined-var}
}

\item[$w_{\rm i}(n)$] SOC error due to current measurement error \eqref{eq:CCerrorSOC_n}
\item[$w_{\rm I}(n)$] SOC error due to current integration error \eqref{eq:INTerrorSOC_n}
\item[$w_{\rm C}(n)$] SOC error due to battery capacity uncertainty \eqref{eq:CBATTerrorSOC_n}
{
\item[$w_\eta(n)$] SOC error due to the uncertainty in c/d efficiency \eqref{eq:CCerrorSOC_eta}
\item[$ w_\Delta (n)$] SOC error due to timing oscillator uncertainty \eqref{eq:TIMEerrorSOC_n}
\item[$ w (n)$] SOC error due combined uncertainties \eqref{eq:soc-err-combined}
}

\item [$z_{\rm i}(k)$] Measured current at time $k$ \eqref{eq:z_i}
{
\item [$z_{\rm v}(k)$] Measured voltage at time $k$ \eqref{eq:meas-model-new}
}

\end{abbreviations}

\section{Problem Definition}
\label{sec:probDef}

The traditional Coulomb counting equation to compute the state of charge (SOC) of a battery at time $t$ is given below \cite{plettBMSbookpart1} 
\begin{eqnarray}
s(t)=s(0) +\frac{\eta }{3600 {\rm C}_{\rm batt}}\int_{0}^{t} i(t)dt
\label{eq:CCct}
\end{eqnarray}
{where 
$\eta$ is the Coulomb counting efficiency defined as 
\begin{align}
\eta = \left\{
\begin{array}{c cc}
\eta_\rc & i(t)>0 & \text{(charging efficiency)} \\
\eta_d & i(t)<0 & \text{(discharging efficiency),}
\end{array}
\right.
\label{eq:ch/dc-efficiency}
\end{align}}
the unit of time $t$ is in seconds,  
$s(t)$ denotes the SOC at time $t$,
$s(0)$ denotes the initial SOC at time $t=0$, 
$i(t)$ is the current in Amperes (A) through the battery at time $t$,
and $C_{\rm batt}$ is the battery capacity in Ampere hours (Ah). 
There are different approaches to compute the initial SOC $s(0)$; the error/uncertainty involved in computing $s(0)$ will remain the same for any value of $t$. In this paper, we do not delve into the error associated with computing $s(0)$ and  assume that $s(0)$ is perfectly known. 

The Coulomb counting equation \eqref{eq:CCct} is written in continuous-time domain. 
Considering that $i(t)$ is not mathematically defined, a discretized Coulomb counting form needs to be adopted in order to perform the integration of  \eqref{eq:CCct}. 
Widely adopted version of the discrete-time, recursive Coulomb counting equation is given below:
\begin{eqnarray}
s(k)=s(k-1) +\frac{\eta }{3600 {\rm C}_{\rm batt}}\int_{t(k-1)}^{t(k)} i(\tau)d\tau
\label{eq:CCcontin}
\end{eqnarray}
where 
$s(k)$ is the SOC of the battery at time $t(k)$,
and
$i(\tau)$ is the measured current at time $\tau$.
By approximating the integration in \eqref{eq:CCcontin} using a rectangular (backward difference) method as
\begin{eqnarray}
\int_{t(k-1)}^{t(k)} i(\tau)d\tau \approx  \Delta_k  i(t(k)) = \Delta_k  i(k)
\label{eq:ccapprox0}
\end{eqnarray}
where $\Delta_k = t(k)-t(k-1)$ is the sampling duration between two adjacent samples.
Now, the widely known form of the Coulomb counting equation can be written as follows \cite{plettBMSbookpart1,plettBMSbookpart2}
\begin{eqnarray}
s(k) = s(k-1) + \frac{\eta \Delta_k  i(k) }{3600 {\rm C}_{\rm batt}} 
\label{eq:CCdiscrete}
\end{eqnarray}

The Coulomb counting equation \eqref{eq:CCdiscrete} is only approximate due to the following sources of errors:
\begin{enumerate}
\item Measurement error in the current $i(k)$ 
\item Error due to the approximation of the integration in \eqref{eq:ccapprox0}
\item Uncertainty in the knowledge of battery capacity ${\rm C}_{\rm batt}$
\item {Uncertainty in the knowledge of the Coulomb counting efficiency $\eta$}
\item The error in the measure of sampling time $\Delta$
\end{enumerate}
In the next four sections of this paper, we mathematically quantify the effect of the above four sources of error in the computed SOC $s(k) $ in \eqref{eq:CCdiscrete}. 
In each section, simulation examples are employed to verify the mathematically derived error quantities.

\section{Individual Uncertainty Analysis}

\subsection{Effect of Current Measurement Error}
\label{sec:CMerror}

The current through the battery is measured using a current sensor that is prone to errors. 
The measured current $z_{\rm i}(k)$ can be modeled as follows 
\begin{eqnarray}
z_{\rm i}(k) = i(k) + n_{\rm i}(k) 
\label{eq:z_i}
\end{eqnarray}
where $i(k)$ is the true current though the battery and $n_{\rm i}(k) $ is the measurement error in the current that can be assumed to be zero-mean with standard deviation $ \sigma_{\rm i}$, i.e.,
\begin{eqnarray}
\begin{aligned}
E\{n_{\rm i}(k) \}  &= 0 \\
E\{n_{\rm i}(k)^2 \} &=  \sigma^2_{\rm i}
\end{aligned}
\label{eq:sig_i}
\end{eqnarray}

Let us substitute the measured current \eqref{eq:z_i} in \eqref{eq:CCdiscrete} and re-write the Coulomb counting equation that considers the current measurement error as follows:
\def\MEASnoise{\rm meas.\,\,  noise}
\def\SOCnoise{\rm SOC \,\, noise}
\def\SOCerror{\rm SOC \,\, error}
\begin{small}
\begin{eqnarray}
s(k+1) &=& s(k) + \frac{\eta\Delta_k  z_{\rm i}(k) }{3600 {\rm C}_{\rm batt}}  \nonumber \\
&=& s(k) + \frac{\eta\Delta_k  i(k) }{3600 {\rm C}_{\rm batt}}  + 
\underbrace{\frac{\eta\Delta_k  n_{\rm i}(k) }{3600 {\rm C}_{\rm batt}}}_{\SOCerror}
\label{eq:CCerror}
\end{eqnarray}
\end{small}
Now, assuming that the sampling time is perfectly known and fixed as   
\begin{eqnarray}
\Delta \triangleq \Delta_k
\label{eq:Delta}
\end{eqnarray}
the SOC at time step $k=1, 2,\ldots$ can be written as,
\begin{eqnarray}
\begin{aligned}
s(0) &=   \text{initial SOC estimation}  \\
s(1) &= s(0) + \frac{\eta\Delta  i(1) }{3600 {\rm C}_{\rm batt}}  + \frac{\eta\Delta  n_{\rm i}(1) }{3600 {\rm C}_{\rm batt}} \\
s(2) &= s(1) + \frac{\eta\Delta  i(2) }{3600 {\rm C}_{\rm batt}}  + \frac{\eta\Delta  n_{\rm i}(2) }{3600 {\rm C}_{\rm batt}} \\
&= s(0) +\frac{\eta\Delta [ i(1)+ i(2)] }{3600 {\rm C}_{\rm batt}}  + \frac{\eta\Delta  [n_{\rm i}(1)+n_{\rm i}(2)] }{3600 {\rm C}_{\rm batt}} 
\end{aligned}
\end{eqnarray}
Considering $n$ consecutive samples, the SOC at time $k=n$ can be shown to be 
\begin{small}
\begin{eqnarray}
s(n) &=& s(0) + 
\underbrace{\frac{\eta\Delta}{3600 {\rm C}_{\rm batt}}{\displaystyle\sum_{k=1}^{n}} i(k) }_{s_{\rm CC}(n)} + 
\underbrace{\frac{\eta\Delta}{3600 {\rm C}_{\rm batt}}{\displaystyle\sum_{k=1}^{n}}n_ i(k)}_{w_{\rm i}(n)} \nonumber\\
&=& s(0) + s_{\rm CC}(n) + w_{\rm i}(n)
\label{eq:CCerrorSOC_n}
\end{eqnarray}
\end{small}
{where, 
$s_{\rm CC}(n)$ indicates the change in SOC from time $k=0$ until $k=n$ and 
$w_{\rm i}(n)$ is the error in the computed SOC at time $k=n$.}

{It can be noticed that the change in SOC can be decomposed into charging Coulombs and discharging SOC as follows 
\begin{align}
s_{\rm CC}(n) = s_{\rm CCc}(n) +  s_{\rm CCd}(n)
\label{eq:SOCdiff}
\end{align}
where 
\begin{align}
s_{\rm CCc}(n) &= \frac{\eta_\rc\Delta}{3600 {\rm C}_{\rm batt}}{\displaystyle\sum_{k=1}^{n}} i(k) \times [i(k) > 0] \\
s_{\rm CCd}(n) &= \frac{\eta_d\Delta}{3600 {\rm C}_{\rm batt}}{\displaystyle\sum_{k=1}^{n}} i(k) \times [i(k) < 0]
\end{align}
where the logical quantity $ [i(k) > 0] $ is defined as
\begin{align}
 [i(k) > 0] =
 \left\{
 \begin{array}{cc}
 1 & i(k)>0\\
 0 & i(k)<0
 \end{array}
 \right.
\end{align}
and the logical quantity $ [i(k) > 0] $ is defined as
\begin{align}
 [i(k) < 0] =
 \left\{
 \begin{array}{cc}
 1 & i(k)<0\\
 0 & i(k)>0
 \end{array}
 \right.
\end{align}
}

{Similarly, the error in the computed computed SOC can be split into two terms corresponding to charging and discharging, i.e., 
\begin{align}
w_{\rm i}(n) = w_{\rm ic}(n) +  w_{\rm id}(n)
\label{eq:wi-c/d}
\end{align}
where
\begin{align}
 w_{\rm ic}(n) &= \frac{\eta_\rc \Delta}{3600 {\rm C}_{\rm batt}}{\displaystyle\sum_{k=1}^{n}}n_ i(k)\times [i(k) > 0] \label{eq:wi-c} \\
 w_{\rm id}(n) &= \frac{\eta_d \Delta}{3600 {\rm C}_{\rm batt}}{\displaystyle\sum_{k=1}^{n}}n_ i(k)\times [i(k) < 0] \label{eq:wi-d}
\end{align}
It must be noted that the current measurement noise $n_\ri(k) \sim \cN(0, \sigma_\ri^2)$ has the same characteristics during charging and discharging. 
 }

Now, it can be verified that the SOC error $w_{\rm i}(n)$ has the following properties
\begin{eqnarray}
\begin{aligned}
E\{w_{\rm i}(n) \}  &= 0 \\
E\{w_{\rm i}(n)^2 \} &= \sigma_{\rm s, i} (n)^2 =   \frac{\Delta^2 \sigma^2_{\rm i} }{3600^2 {\rm C}_{\rm batt}^2} ( \eta_\rc n_\rc + \eta_\rd n_\rd)
\end{aligned}
\label{eq:STD-imeas-error}
\end{eqnarray}
{
where 
$n_\rc$ is the number of current charging current samples and 
$n_\rd$ is the number of current discharging current samples that satisfy
\begin{align}
n_\rc + n_\rd = n
\end{align}
}
It can be noted that as $n \rightarrow \infty $, the noise variance of the computed SOC error also approaches infinity.  
Let us write the SOC noise due to current measurement error in a simplified form as follows:
\begin{align}
\sigma_{\rm s,i}(n) 
&= \left(  \frac{\Delta  \rho_{\rm i}}{3600  }  \right)  \sqrt{{\eta_\rc n_\rc + \eta_\rd n_\rd}}
\label{eq:SOCerr/hour}
\end{align}
where the ratio between the measurement noise standard deviation and battery capacity (in Ah), denoted in this paper as the {\em current measurement noise coefficient} (which has a unit of ${\rm hour}^{-1}$), is defined as
\begin{align}
{\rm \rho_i} = \frac{\rm \sigma_i}{\rm C_{batt}}
\label{eq:rho_i}
\end{align}
It must be noted that since the SOC $s(n)$ is defined within $[0,1]$. However, SOC is usually displayed in percentage. 
As such, the standard deviation of the SOC error in \eqref{eq:SOCerr/hour} is given in percentage as follows:
\begin{align}
\sigma_{\rm s,i}(n) \quad \text{in \%}
&=  \left(  \frac{\Delta  \rho_{\rm i}}{36  }  \right)  \sqrt{{\eta_\rc n_\rc + \eta_\rd n_\rd}} \quad \%
\label{eq:SOC-Err-SD}
\end{align}

Table \ref{table:SOCvarMeasErrorindevidual} shows the standard deviation (s.d.) in the SOC error due to current measurement error for different sampling intervals over different durations of time under the above assumptions. 
Here it is assumed that the battery capacity is ${\rm C_{batt}} = 1.5 \,\, {\rm Ah}$ and the current measurement error standard deviation is $\sigma_{\rm i} = 10 \,\, {\rm mA.}$
\begin{table}[h]
\caption{SOC error s.d. (\%) due to current measurement error}
\label{table:SOCvarMeasErrorindevidual}
\begin{center}
\begin{tabular}{|l|l|l|l|}
\hline
& 1 hour & 24 hours  & 1 year   \\ 
\hline
$\Delta = 0.1 \, {\rm s}$ & 0.0035  &  0.0172   & 0.3289  \\
\hline
$\Delta = 1 \, {\rm s}$ & 0.0111 &   0.0544   &  1.0399   \\
\hline
$\Delta = 10 \, {\rm s}$ & 0.0351   &   0.1721   &  3.2886  \\
\hline
\end{tabular}
\end{center}
\end{table}

It must be noted that the SOC error shown in Table \ref{table:SOCvarMeasErrorindevidual} is computed assuming zero uncertainties in all the other sources of error (integration, capacity, timing oscillator) and the initial SOC $s(0)$. 

The variance of the SOC error \eqref{eq:SOC-Err-SD} due to current measurement error keeps increasing with time. 
As such, we denote this as a {\em time-cumulative} error. 
For time-cumulative errors, the standard deviation of the error keeps increasing with time -- if it is not reset, it will completely corrupt the estimated SOC. 
A possible approach to reduce time-cumulaitive error is by resetting the Coulomb count to $s(k) = s(0)$ once in a while. 
Considering that the reset value of SOC also comes with errors (that is not considered in this paper) it is important to select an instant where the uncertainty in the reset SOC will be smaller than the uncertainty derived in \eqref{eq:SOC-Err-SD}.

%\clearpage
\subsection{Effect of Approximating Current Integration }
\label{sec:CIerror}

\begin{figure}[htbp]
\begin{center}
\includegraphics[width = .75\columnwidth]{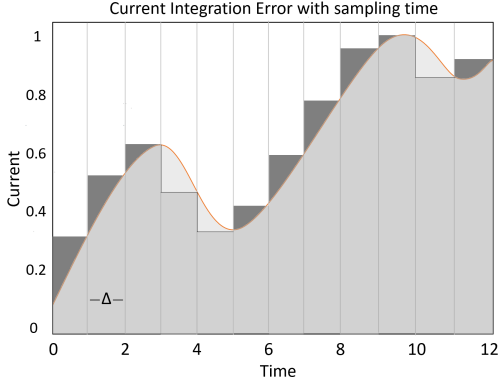}
\caption{
{\bf A generic illustration of the current integration error.}
The integration error $\delta_{\rm I}(k)$ is shown in shade. 
It can be noticed that the integration error can be both positive and negative -- the dark shade indicates positive error and the light shade indicates negative error. 
Based on this observation, the integration error is assumed to be zero-mean. 
}
\label{fig:SOCintsamplerror}
\end{center}
\end{figure}

The Coulomb counting approach summarized in the previous section approximates the integration of current over time using a simple first-order (rectangular) approximation (see \eqref{eq:ccapprox0}).
A generic rectangular approximation to integration is illustrated in Figure \ref{fig:SOCintsamplerror}. 
For such rectangular approximation, the {\em integration error} $ \delta_{\rm I}(k)$ is defined as the difference between the true integral and the approximation, i.e.,
\begin{eqnarray}
\underbrace{\int_{\tau(k-1)}^{\tau(k)} i(\tau)d\tau}_{\text{true integration}} = 
\underbrace{\Delta  i(k)}_{\text{approximation}} + \,\,\,\,  \underbrace{\delta_{\rm I}(k)}_{\text{integration error}}
\label{eq:int_error}
\end{eqnarray}
\def\INTerror{\rm Integ. Error}

The nature of the integration error $\delta_{\rm I}(k)$ is of specific interest. 
It can be observed that, for rectangular approximation, the integration error is proportional to the sampling duration $\Delta$ \cite{chapra1998numerical}, i.e.,
\begin{align}
\delta_{\rm I}(k) &\propto \Delta 
\label{eq:int_propto_del}
\end{align}
Further, the integration error is proportional to the difference in the adjacent samples of measured current, i.e., 
\begin{align}
\delta_{\rm I}(k) &\propto \left[ i(k)- i(k-1) \right] 
\label{eq:int_propto_id}
\end{align}
Since, $ [ i(k)- i(k-1)] $ in \eqref{eq:int_propto_id} is a time varying quantity, we can approximately write
\begin{align}
\delta_{\rm I}(k) &\propto  \sigma_{\rm L} 
\label{eq:int_propto_sigL}
\end{align}
where $ \sigma_{\rm L} $ is the standard deviation of the load (or charging) current (e.g., if the current is constant then $ \sigma_{\rm L} =0$ and so is the integration error). 
In addition, the sign of the integration error is both positive and negative when there is variance in the magnitude of the current $i(k)$ -- see Figure \ref{fig:SOCintsamplerror} for an illustration of this.  Using this observation, we can write 
\begin{align}
E\{\delta_{\rm I}(k) \}  \approx 0
\end{align}
That is, considering a large number of samples, we can assume the error due to the rectangular approximation of current-integration to be zero-mean.

Based on the discussion so far, the integration error has the following (approximate) properties. 
\begin{eqnarray}
\begin{aligned}
E\{\delta_{\rm I}(k) \}  &= 0 \\
E\{\delta_{\rm I}(k) ^2 \} &= \sigma^2_{\rm I}
\end{aligned}
\end{eqnarray}
where $ \sigma^2_{\rm I}$ is the variance of the current integration error. 
From \eqref{eq:int_propto_del} and \eqref{eq:int_propto_sigL}, we can write 
\begin{eqnarray}
\begin{aligned}
\sigma_{\rm I} & \propto \Delta \sigma_{\rm L}  \\
&= \kappa  \Delta \sigma_{\rm L}
\label{eq:int_err_sigI}
\end{aligned}
\end{eqnarray}
where 
$ \kappa$ is a constant, 
$\Delta $ is the sampling time, and
$\sigma_{\rm L} $ is the standard deviation of the load current. 

\begin{figure}[htbp]
\begin{center}
\subfloat[][Smart Phone \cite{avvari2015experimental}, $\sigma_{\rm L} = 0.1673 \, {\rm A}, {\rm C}_{\rm batt} = 1.5 {\rm Ah}$]
{\includegraphics[width = .75\columnwidth]{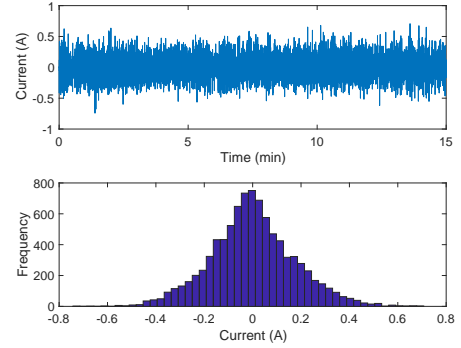}}\\
\subfloat[][Electric Vehicle  \cite{plettBMSbookpart2}, $\sigma_{\rm L} =  8.6917 \, {\rm A},  {\rm C}_{\rm batt} \approx 250 {\rm Ah}$]
{\includegraphics[width = .75\columnwidth]{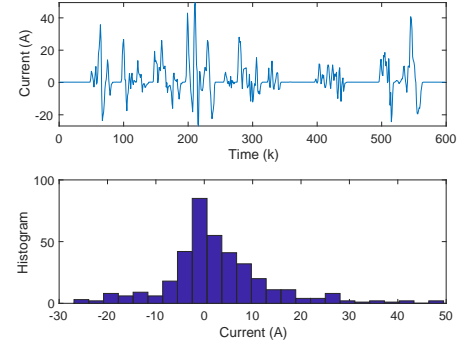}}
\caption{
{\bf Current difference in realistic loads.}
In both (a) and (b), the top plot shows the current difference $[i(k)-i(k-1)]$ in typical load profile in Amperes and the plot at the bottom shows the magnitude of the current difference as a histogram.
%\blue{y-axis is $i(k)-i(k-1)$ not $i(k)$ --- need to fix this.}***
}
\label{fig:RealLoad}
\end{center}
\end{figure}

Figure \ref{fig:RealLoad} shows two different load current profiles from practical applications. 
It supports the assumption made in \eqref{eq:int_propto_id} that the current difference [$i(k)-i(k-1)]$ indeed is zero mean. 

By following the same approach of Section \ref{sec:CMerror}, we can write the computed SOC in recursive form as
\begin{eqnarray}
s(k+1) 
&=& s(k) + \frac{\eta(\Delta i(k)+\delta_{\rm I} (k) )}{3600 {\rm C}_{\rm batt}}  \nonumber \\
&=& s(k) + \frac{\eta\Delta  i(k) }{3600 {\rm C}_{\rm batt}}  + 
\underbrace{\frac{\eta  \delta_{\rm I}(k) }{3600 {\rm C}_{\rm batt}} }_{\INTerror}
\label{eq:CCinterror}
\end{eqnarray}
where the integration error is incorporated based on \eqref{eq:int_error}. 

Now, let us write the SOC at time step $k=0, 1, 2 \ldots$ as
\begin{eqnarray}
s(0) &=&  \text{initial SOC estimation} \nonumber \\
s(1) &=& s(0) + \frac{\eta\Delta i(1) }{3600 {\rm C}_{\rm batt}}  + \frac{\eta \delta_{\rm I}(1) }{3600 {\rm C}_{\rm batt}} \\
s(2) &=& s(1) + \frac{\eta\Delta i(2) }{3600 {\rm C}_{\rm batt}}  + \frac{\eta \delta_{\rm I}(2) }{3600 {\rm C}_{\rm batt}} \nonumber\\
&=& s(0) +\frac{\eta\Delta [ i(1)+ i(2)] }{3600 {\rm C}_{\rm batt}}  + \frac{\eta(\delta_{\rm I}(1)+\delta_{\rm I}(2)) }{3600 {\rm C}_{\rm batt}} 
\end{eqnarray}
Considering $n$ consecutive samples, the computed SOC at time $k=n$, can be shown to be 
\begin{small}
\begin{eqnarray}
s(n) &=& s(0) + \underbrace{\frac{\eta\Delta}{3600 {\rm C}_{\rm batt}}{\displaystyle\sum_{k=1}^{n}} i(k)}_{s_{\rm CC}(n)}  +\underbrace{ \frac{\eta}{3600 {\rm C}_{\rm batt}}{\displaystyle\sum_{k=1}^{n}}\delta_{\rm I}(k) }_{w_{\rm I} (n)}\nonumber\\
&=& s(0) + s_{\rm CC}(n) + w_{\rm I}(n)
\label{eq:INTerrorSOC_n}
\end{eqnarray}
\end{small}
where $w_{\rm I}(n)$ is SOC error due to the approximation of integration.  

{
Similar to \eqref{eq:wi-c/d}--\eqref{eq:wi-d}, the SOC error $w_{\rm I}(n)$ can be decomposed, corresponding to charging and discharging, as follows 
\begin{align}
w_{\rm I}(n) = w_{\rm Ic}(n) +  w_{\rm Id}(n)
\label{eq:wI-c/d}
\end{align}
where
\begin{align}
 w_{\rm Ic}(n) &= \frac{\eta_\rc }{3600 {\rm C}_{\rm batt}}{\displaystyle\sum_{k=1}^{n}} \delta_{\rm I} (k)\times [i(k) > 0] \label{eq:wI-c} \\
 w_{\rm Id}(n) &= \frac{\eta_d }{3600 {\rm C}_{\rm batt}}{\displaystyle\sum_{k=1}^{n}} \delta_{\rm I} (k)\times [i(k) < 0] \label{eq:wI-d}
\end{align}
}

It can be noted that the SOC error due to integration has the following properties
\begin{eqnarray}
\begin{aligned}
E\{w_{\rm I}(n) \}  &= 0 \\
E\{w_{\rm I}(n)^2 \} &= \sigma_{\rm s,I}(n)^2 =   \frac{\kappa^2 \Delta^2  \sigma^2_{\rm L} }{3600^2 {\rm C}_{\rm batt}^2} {( \eta_\rc n_\rc + \eta_\rd n_\rd )} 
\end{aligned}
\label{eq:STD-integ-error}
\end{eqnarray}
The standard deviation of integration error is 
\begin{eqnarray}
\sigma_{\rm s,I}(n) 
&=&  \frac{\kappa \Delta  \rho_{\rm I} }{3600 } \sqrt{{\eta_\rc n_\rc + \eta_\rd n_\rd}} 
\label{eq:sigmainteq}
\end{eqnarray}
where the {\em integration error coefficient} is defined as 
\begin{align}
\rho_{\rm I} = \frac{\rm \sigma_L}{\rm C_{batt}}
\label{eq:rho_I}
\end{align}

Considering that the SOC $s(n)$ is defined within $[0,1]$, the standard deviation of the SOC in \eqref{eq:STD-integ-error} ranges between 
$\sigma_{\rm s,I}(n) \in [0, 1]$.
Usually, SOC is displayed in percentage. 
As such, the standard deviation of the SOC error in \eqref{eq:sigmainteq} can be displayed in percentage as follows 
\begin{align}
\sigma_{\rm s,I}(n) /\text{year (in \%)} 
&=  \frac{\kappa \Delta  \rho_{\rm I} }{36 } \sqrt{{\eta_\rc n_\rc + \eta_\rd n_\rd}} \quad \%
\label{eq:SOC-Err-SD-IntErr}
\end{align}

Now, let us make some realistic assumptions in order to simplify the above expression further. 
Based on the data shown in Figure \ref{fig:RealLoad},  we have 
\begin{align}
\rho_{\rm I} =
\left\{
\begin{array}{cl}
0.1115 & \text{Smart Phone} \\
0.0348 & \text{Electric Vehicle}
\end{array}
\right. 
\label{eq:rho_values}
\end{align}

Table \ref{table:IntSOCerr_SmPh} and Table \ref{table:IntSOCerr_EV} show the computed SOC error standard deviation 
due to the {\em current integration error} for different sampling intervals over longer periods of time.
These two tables are made based on the values shown in \eqref{eq:rho_values} and by assuming $\kappa = 1$. 
\begin{table}[h]
\caption{S.D. of SOC Error (\%) - Smart Phone Data}
\label{table:IntSOCerr_SmPh}
\begin{center}
\begin{tabular}{|l|l|l|l|l|}
\hline
& 1 hour & 24 hours  & 1 year  \\ 
\hline
$\Delta = 0.1 \, {\rm s}$ &  0.0588   & 0.2879  &  5.5002    \\
\hline
$\Delta = 1 \, {\rm s}$ & 0.1858   &  0.9104  &  17.3930    \\
\hline
$\Delta = 10 \, {\rm s}$ & 0.5877   &  2.8789  &  55.0016   \\
\hline
\end{tabular}
\end{center}
\end{table}

\begin{table}[h]
\caption{S.D. of SOC Error (\%) - EV Data}
\label{table:IntSOCerr_EV}
\begin{center}
\begin{tabular}{|l|l|l|l|l|}
\hline
& 1 hour & 24 hours  & 1 year  \\ 
\hline
$\Delta = 0.1 \, {\rm s}$ &0.0183   & 0.0899  &  1.7166\\
\hline
$\Delta = 1 \, {\rm s}$ &  0.0580  &  0.2841  &  5.4285 \\
\hline
$\Delta = 10 \, {\rm s}$ & 0.1834   &  0.8985  & 17.1664 \\
\hline
\end{tabular}
\end{center}
\end{table}

%\clearpage
\subsection{Effect of the Uncertainty in Battery Capacity}
\label{sec:CbattError}

Battery capacity is the amount of coulombs that can be charged to (or discharged from) the battery. 
The battery capacity fades over  time \cite{barre2013review} and the rate of capacity fade depends on calendar life as well as environmental and usage patterns the battery has experienced over long periods of time \cite{spotnitz2003simulation}.
Thus, true value of the battery capacity ${\rm C_{batt}}$ is not precisely known.
Usually {\em a measure} of the battery capacity, denoted ${\rm C_{batt}}$,  is used to estimate the battery SOC. 
Such a capacity measure is not exact and it relates to the true battery capacity as follows
\begin{eqnarray}
 {\rm C_{batt}}  = {\rm C_{true}} +  {\rm C}_{\Delta}  
 \label{eq:cbatt-ctrue}
\end{eqnarray}
where ${\rm C}_{\Delta} $ represents the uncertainty in the knowledge about the true battery capacity $ {\rm C_{true}}  $.
For instance, it was argued in \cite{balasingam2014robustP2} that this uncertainty can be modeled as a zero-mean Gaussian distribution, i.e.,
\begin{eqnarray}
{\rm C}_{\Delta}   \sim \cN(0, \sigma^2_{\rm batt} )
\label{eq:CbattError}
\end{eqnarray}
where $\sigma_{\rm batt}$ is the standard deviation of the capacity estimation error.

The first order Taylor series approximation of a function $f(x)$ around a point $x_0$ is given by
\begin{eqnarray}
f(x) = f(x_0) + (x-x_0)\Delta f'(x_0)
\end{eqnarray}
using the above Taylor series approximation and the relationship \eqref{eq:cbatt-ctrue} the inverse capacity can be approximated as follows
\begin{eqnarray}
\frac{1}{{\rm C}_{\rm batt}} \approx \frac{1}{{\rm C}_{\rm true}} - \frac{{\rm C}_\Delta}{{\rm C}_{\rm true}^2}
\label{eq:TaylorApprox}
\end{eqnarray}
\def\BATerror{\rm Cbatt.  Error}

With the above approximation to the inverse capacity, let us re-write the Coulomb counting equation as follows
\begin{eqnarray}
s(k+1) &=& s(k) + \frac{\eta \Delta i(k) }{3600 {\rm C}_{\rm batt}}  \nonumber \\
&=& s(k) + \left( \frac{\eta \Delta  i(k) }{3600 } \right)\times \left(\frac{1}{{\rm C}_{\rm true}}  - \frac{\rm C_\Delta}{{\rm C}_{\rm true}^2} \right)  \nonumber \\
&=& s(k) + \frac{\eta \Delta  i(k) }{3600 {\rm C}_{\rm true}} -   \frac{\eta \Delta  i(k){\rm C_\Delta} }{3600 {\rm C}_{\rm true}^2}
\label{eq:CC_cbatt_err}
\end{eqnarray}
Now, SOC at time step $k= 0, 1,2,\ldots$ can be written as
\begin{eqnarray}
s(0) &=&  \text{initial SOC estimation} \nonumber \\
s(1) &=& s(0) + \frac{\eta \Delta  i(1) }{3600 {\rm C}_{\rm true}}  - \frac{\eta  \Delta  i(1){\rm C_\Delta} }{3600 {\rm C}_{\rm true}^2} \\
s(2) &=& s(1) + \frac{\eta \Delta  i(2) }{3600 {\rm C}_{\rm true}}  - \frac{\eta  \Delta  i(2){\rm C_\Delta}}{3600 {\rm C}_{\rm true}^2} \nonumber\\
&=& s(0) +\frac{\eta \Delta [ i(1)+ i(2)] }{3600 {\rm C}_{\rm true}}  - \frac{{\rm C_\Delta}\eta  \Delta [ i(1)+ i(2)] }{3600 {\rm C}_{\rm true}^2}
\end{eqnarray}
Considering $n$ consecutive samples the computed SOC at time $k=n$, can be shown to be 
\begin{align}
s(n) &= s(0) + \underbrace{\frac{\eta \Delta}{3600 {\rm C}_{\rm true}}{\displaystyle\sum_{k=1}^{n}} i(k)}_{s_{\rm CC}(n)}   \underbrace{- \frac{\eta \Delta {\rm C_\Delta}}{3600 {\rm C}_{\rm true}^2}{\displaystyle\sum_{k=1}^{n}} i(k)}_{w_{\rm C} (n)}\nonumber\\
&= s(0) + s_{\rm CC}(n) + w_{\rm C}(n)
\label{eq:CBATTerrorSOC_n}
\end{align}
where $w_{\rm C} (n)$ is the SOC error due to the uncertainty in battery capacity.
{
Similar to \eqref{eq:wi-c/d}--\eqref{eq:wi-d}, $w_{\rm C}(n)$ above can be decomposed into the following two terms:
\begin{align}
w_{\rm C}(n) = w_{\rm Cc}(n) +  w_{\rm Cd}(n)
\label{eq:wi-c/d}
\end{align}
where
\begin{align}
 w_{\rm Cc}(n) &= \frac{\eta_\rc \Delta {\rm C_\Delta} }{3600 {\rm C}_{\rm true}^2}{\displaystyle\sum_{k=1}^{n}} i(k)\times [i(k) > 0] \label{eq:wi-c} \\
 w_{\rm Cd}(n) &= \frac{\eta_d \Delta {\rm C_\Delta} }{3600 {\rm C}_{\rm true}^2}{\displaystyle\sum_{k=1}^{n}} i(k)\times [i(k) < 0] \label{eq:wi-d} \\
\end{align}
Now, $w_{\rm C}(n)$ becomes
\begin{align}
w_{\rm C}(n) &= \left( \frac{{\rm C_\Delta}}{\rm C_{true}} \right)\left( s_{\rm CCc}(n) + s_{\rm CCd}(n) \right) \nonumber \\
 &= \left( \frac{{\rm C_\Delta}}{\rm C_{true}} \right) s_{\rm CC}(n) 
\end{align}
}

Now, we can write the following about the SOC error $w_{\rm C}(n)$ due to the uncertainty in the knowledge of the battery capacity 
\begin{align}
E\{w_{\rm C}(n) \}  &= 0 \\
E\{w_{\rm C}(n)^2 \} &= \sigma_{\rm s,C}(n)^2 =   \frac{E\{ C_{\Delta}^2 \} }{\rm C_{true}^2}  s_{\rm CC}(n)^2  \label{eq:STD-Cbatt-error} \\
&=  \frac{\sigma_{\rm batt}^2 }{\rm C_{true}^2}  s_{\rm CC}(n)^2 = \rho_{\rm C}^2 s_{\rm CC}(n)^2
\end{align}
where the dimensionless {\em capacity uncertainty coefficient} is defined as 
\begin{eqnarray}
 \rho_{\rm C}=  \frac{ \sigma_{\rm batt}} {\rm C_{true}} 
 \label{eq:rhoC}
\end{eqnarray}

%\clearpage
{
\subsection{Effect of the Uncertainty in Charging Efficiency}
\label{sec:TimingError}
Let us assume the uncertainty in charging efficiency as follows 
\begin{align}
\eta_\rc &=  \eta_{\rc \rt}  + \eta_{\rc \Delta} \\
\eta_\rd &=  \eta_{\rd \rt}  + \eta_{\rd \Delta} 
\end{align}
In summary, we may write 
\begin{align}
\eta &=  \eta_{\rt}  + \eta_{\Delta}
\end{align}
where 
\begin{align}
\eta_\rt = \left\{
\begin{array}{cc}
\eta_{\rc \rt} & \text{if} \,\, i(k)>0 \\
\eta_{\rd \rt} & \text{if} \,\, i(k)<0
\end{array}
\right.
\eta_\Delta = \left\{
\begin{array}{cc}
\eta_{\rc \Delta} & \text{if} \,\, i(k)>0 \\
\eta_{\rd \Delta} & \text{if} \,\, i(k)<0
\end{array}
\right.
\end{align}
}

{
Let us substitute the measured current \eqref{eq:z_i} in \eqref{eq:CCdiscrete} and re-write the Coulomb counting equation that considers the current measurement error as follows:
\def\MEASnoise{\rm meas.\,\,  noise}
\def\SOCnoise{\rm SOC \,\, noise}
\def\SOCerror{\rm SOC \,\, error}
\begin{small}
\begin{align}
s(k+1) 
&= s(k) + \frac{\eta_\rt \Delta  i(k) }{3600 {\rm C}_{\rm batt}}  + 
{\frac{\eta_\Delta \Delta   i(k) }{3600 {\rm C}_{\rm batt}}}
\label{eq:CCerror}
\end{align}
\end{small}
Considering $n$ consecutive samples, the SOC at time $k=n$ can be shown to be 
\begin{small}
\begin{align}
s(n) &= s(0) + 
\underbrace{\frac{\eta_\rt \Delta}{3600 {\rm C}_{\rm batt}}{\displaystyle\sum_{k=1}^{n}} i(k) }_{s_{\rm CC}(n)} + 
\underbrace{\frac{\eta_\Delta \Delta}{3600 {\rm C}_{\rm batt}}{\displaystyle\sum_{k=1}^{n}} i(k)}_{w_\eta(n)} \nonumber\\
&= s(0) + s_{\rm CC}(n) + w_\eta(n)
\label{eq:CCerrorSOC_eta}
\end{align}
\end{small}
where the SOC error $w_\eta(n)$ can be expressed as the following 
\begin{align}
w_\eta(n) = & \frac{\eta_{\rc \Delta}}{\eta_{\rc \rt}} \left(   \frac{ \eta_{\rc \rt} \Delta }{3600 {\rm C}_{\rm batt}} {\displaystyle\sum_{k=1}^{n}} i(k)\times [i(k) > 0] \right) \nonumber \\
& + \frac{\eta_{\rd \Delta}}{\eta_{\rd\rt} } \left(   \frac{ \eta_{\rd\rt} \Delta }{3600 {\rm C}_{\rm batt}} {\displaystyle\sum_{k=1}^{n}} i(k)\times [i(k) < 0] \right) \nonumber \\
 = & \frac{\eta_{\rc \Delta}}{\eta_{\rc\rt}} s_{\rm CCc}(n)  + \frac{\eta_{\rd \Delta}}{\eta_{\rd\rt}} s_{\rm CCd}(n) \nonumber \\
  = & \rho_{\eta_\rc} s_{\rm CCc}(n)  + \rho_{\eta_\rd} s_{\rm CCd}(n)
\end{align}
where
\begin{align}
\rho_{\eta_\rc} =   \frac{\eta_{\rc \Delta}}{\eta_{\rc\rt}} \quad \text{and} \quad \rho_{\eta_\rd} =   \frac{\eta_{\rd \Delta}}{\eta_{\rd\rt}}
\label{eq:rho-charging-efficiency}
\end{align}
are defined as the 
{\em charging uncertainty coefficient} and the {\em discharging uncertainty coefficient}, respectively.
Let us model these two coefficients as  
$\rho_{\eta_\rc} \sim \cN(0, \sigma_{\eta_c}^2)$ and
$\rho_{\eta_\rd} \sim \cN(0, \sigma_{\eta_d}^2)$.
With this assumption, it can be shown that the SOC error $w_\eta (n)$ has the following properties
\begin{align}
E\{w_\eta(n) \}  &= 0 \\
E\{w_\eta (n)^2 \} &= \sigma_{\rm s, \eta} (n)^2 =  \sigma_{\eta_c}^2 s_{\rm CCc}(n)^2   + \sigma_{\eta_d}^2 s_{\rm CCd}(n)^2  
\label{eq:STD-c/d-coeff}
\end{align}
}

\subsection{Effect of the Uncertainty in Timing Oscillator}
\label{sec:TimingError}

%For this source of uncertainty, we have the time oscillator that they're utilizing a crystal with certain frequency\cite{vig1992introduction}. 
%The accuracy of a time oscillator define as part per minute (PPM) which considering that by crystal frequency, we can come up with the uncertainty that we have for time \cite{vig1992introduction}. 
%In addition to that, each crystal based oscillator have a capacitive load and the accuracy of the time oscillator is based on the accuracy of oscillator and the differentiation between the capacitive load and the crystal. \cite{crystalvec}.

The timing oscillator 
Hence, for this approach we have
\begin{eqnarray}
\Delta =  \Delta_{\rm true} + \Delta_{\epsilon}
\label{time:error}
\end{eqnarray}
where $\Delta_\epsilon$ is the timing oscillator error which is not a random parameter.
The timing oscillator error $\Delta_\epsilon$ acts like a bias --- we consider it to be a constant over long periods of time. 
Also, let us quantify the {\em timing error coefficient} as follows 
\begin{eqnarray}
\rho_\Delta = \frac{\Delta_\epsilon} { \Delta_{\rm true}}
\label{eq:TEcoefficient}
\end{eqnarray}
Let us assume that a timing oscillator is off by three minutes in one month (30 days); in this case the constant $\rho_\Delta $ will be
\begin{eqnarray}
\rho_\Delta  = \frac{3}{30 \times 24 \times 60} = 6.9444 \times 10^{-5} \approx 69\times 10^{-6}
\label{eq:eta}
\end{eqnarray}

Using \eqref{time:error} in main Coulomb counting equation \eqref{eq:CCdiscrete} we have
\def\timerror{\rm Time. Error}
\begin{small}
\begin{eqnarray}
s(k+1) 
&=& s(k) + \frac{\eta \Delta }{3600 {\rm C}_{\rm batt}} \sum_{k=1}^{n}  i(k)   \nonumber \\
&=& s(k) + \frac{\eta \Delta_{\rm true} }{3600 {\rm C}_{\rm batt}} \sum_{k=1}^{n}  i(k)  
+ \frac{\eta \Delta_\epsilon }{3600 {\rm C}_{\rm batt}} \sum_{k=1}^{n}  i(k)   \nonumber \\
&=& s(0) + s_{\rm CC}(n) + w_\Delta (n)
\label{eq:TIMEerrorSOC_n}
\end{eqnarray}
\end{small}
The SOC estimation error can be simplified as 
\begin{eqnarray}
w_{\rm \Delta} (n) &=&  \left( \frac{\eta}{3600 {\rm C_{batt}} } \sum_{k=1}^{n} i(k) \Delta_\epsilon \right)  \nonumber \\
&=&  \rho_\Delta  \left( \frac{\eta}{3600 {\rm C_{batt}} } \sum_{k=1}^{n} i(k) \Delta_{\rm true} \right) \nonumber  \\
&=&  \rho_\Delta  s_{\rm CC}(n)
\label{eq:STD-tOSC-error}
\end{eqnarray}
Assuming that the initial SOC $s(0)$ is zero, it can be said that 
\begin{eqnarray}
0 \leq s_{\rm CC}(n) \leq 1
\end{eqnarray}
Hence, the SOC error varies between 
\begin{eqnarray}
0 \leq w_{\rm \Delta} (n)  \leq \rho_\Delta 
\end{eqnarray}

{
The SOC error $w_{\rm \Delta} $ is a deterministic quantity for a given battery provided that $\rho_\Delta$ is known. 
However, a realistic assumption is that the knowledge of $\rho_\Delta$ is only probabilistic.
Let us assume that $\rho_\Delta \sim \cN(0, \sigma_\Delta^2)$.
Under this scenario, the SOC error $w_{\rm \Delta} $ has the following properties 
\begin{align}
E\{w_\Delta(n) \}  &= 0 \\
E\{w_\Delta(n)^2 \} &= \sigma_{\rm s, \Delta} (n)^2 =   \sigma_\Delta^2 s_{\rm CC}(n)^2    
\label{eq:err-var-del}
\end{align}
}

Considering that $\rho_\Delta  $ is a very small number, see \eqref{eq:eta}, the error in SOC due to timing oscillator error can be considered to be negligible.

%\clearpage
\section{Summary of Individual Errors}
\label{sec:summary}

In this paper, we present a critical look at Coulomb counting method that is employed to estimate the state of charge of a battery. 
The Coulomb counting approach computes the present SOC as 
\begin{eqnarray}
s(t)= \underbrace{s(0)}_{\text{initial SOC}} + \underbrace{ \int_{0}^{t}\frac{i(\tau) }{ 3600 {\rm C}_{\rm batt}} d\tau}_{\text{change in SOC}}
\nonumber
\end{eqnarray}
where $i(t)$ is the instantaneous current through the battery and ${\rm C}_{\rm batt}$ is the battery capacity in Ampere hours. 
That is, the present SOC is the summation of initial SOC and the change in SOC that is computed through the above integration.  
The SOC can be approximately computed in a recursive manner as follows 
\begin{align}
s(n) &= s(0) + \frac{\Delta  }{3600 {\rm C}_{\rm batt}} \sum_{k=1}^{n} i(k) \nonumber \\
&= \underbrace{s(0)}_{\text{initial SOC}} +  \underbrace{s_{\rm CC}(n)}_{\text{change in SOC}} \nonumber
\end{align}
where $s(k)$ denotes the SOC at time instance $k$,
$ i(k)$ is the measured current at time instance $k$,
and $\Delta$ is the sampling time in seconds. 
That is, the SOC at time $n$ is the summation of the initial SOC $s(0)$ and the accumulated SOC $s_{\rm CC}$ from time $n=0$ until $n$.

In this paper, we showed that the above (discrete) recursive approximation to computing SOC suffers from four sources of error: current measurement error, current integration error, battery capacity uncertainty and the timing oscillator error. 
Particularly, we computed the exact amount of the resulting SOC uncertainty as a result of the above four types of errors. Those results are 
\begin{enumerate}
\renewcommand{\theenumi}{\Alph{enumi}}
\item {\em Current measurement error:}
Considering that the current measurement error is zero-mean with standard deviation $\sigma_{\rm i},$ the computed SOC at time $n$ can be written as 
\begin{align}
s(n) =s(0) + s_{\rm CC}(n) + w_{\rm i}(n) \nonumber
\end{align}
where 
$s(0)$ is the initial SOC and 
$s_{\rm CC}(n)$ is the accumulated SOC from the start at $n=0$.
The SOC error $w_{\rm i}(n)$ is shown to be zero mean with standard deviation (see \eqref{eq:STD-imeas-error})
\begin{eqnarray}
\sigma_{\rm s, i}(n) =   \left(  \frac{\Delta  \rho_{\rm i}}{36  }  \right)  \sqrt{n} \quad \%  
\label{eq:sig_s_i_concl}
\end{eqnarray}
It must be noted that the variance of the Coulomb counting error due to current measurement noise is {\em accumulative with time}. 
As the time increases, i.e., $n \rightarrow \infty$, so does the standard deviation of the SOC error.

\item {\em Current integration error:}
Considering that the current integration is approximated using a rectangular method, the resulting approximation error is shown to be zero-mean with standard deviation $\sigma_{\rm I}$. 
As a result, the computed SOC at time $n$ can be written as 
\begin{eqnarray*}
s(n) = s(0) + s_{\rm CC}(n) + w_{\rm I}(n)
\end{eqnarray*}
where the SOC error $w_{\rm I}(n)$ is shown to be zero mean with standard deviation
\begin{eqnarray}
\sigma_{\rm s,I}(n) =   \frac{\kappa \Delta  \rho_{\rm I} }{36 } \sqrt{n} \quad \%  
\label{eq:sig_s_I_concl}
\end{eqnarray}
Once again, it can be noticed that the variance of the Coulomb counting error due to current integration approximation error is accumulative with time.

\item {\em Uncertainty in the knowledge of battery capacity:}
Considering that the uncertainty in the knowledge of battery capacity is zero-mean with standard deviation $\sigma^2_{\rm batt}$, the SOC at time $n$ is derived as  
\begin{align*}
s(n) = s(0) + s_{\rm CC}(n) + w_{\rm C}(n)
\end{align*}
where the SOC error $w_{\rm C}(n)$ is shown to be zero mean with standard deviation
\begin{align*}
\sigma_{\rm s,C}(n)^2 =   \rho_{\rm C}^2 s_{\rm CC}(n)^2
\end{align*}
where $ \rho_{\rm C}$ is defined as the capacity uncertainty coefficient.
It must be noted that the variance of the capacity uncertainty error is {\em not} accumulative with time, rather, it is proportional to the accumulated SOC $s_{\rm CC}(n) \in [0, 1].$ 
In other words the SOC error due to uncertainty in the knowledge of battery capacity, $w_{\rm C}(n)$, alternates between zero and $\rho_{\rm C}.$

However, depending on the value of $ \rho_{\rm C}$ (the ratio between the s.d. of the uncertainty and the assumed battery Capacity ${\rm C_{batt}}$) the error could be anywhere between zero and 100\%. 
For example, let us assume that $\rho_{\rm C} = 0.1$ and let us assume that the computed SOC at time $n$ is $s(n) = 40\%$. The standard deviation of the uncertainty in the computed $s(n)$ is $0.1 s(n) = 0.1 \times 40 = 4 \%.$
That is, the true SOC can be anywhere between $32\%$ and $48 \%$ with $95\%$ confidence. 
This can be extended to different levels of confidence as follows:
\begin{center}
\begin{tabular}{|c|c|}
\hline
{\bf Where true SOC is?} & {\bf Confidence}\\
\hline
$36\% - 44 \%$ & 68 \% \\
\hline
$32\% -48 \%$ & 95 \% \\
\hline
$28\% -52 \%$ & 99.7 \% \\
\hline
\end{tabular}
\end{center}

{
\item {\em Charging efficiency error:}
The charging and discharging efficiencies are denoted $\eta_c$ and $\eta_d$, respectively.  
The uncertainties in charging and discharging efficiencies are denoted $\eta_{\rc \Delta}$ and $\eta_{\rd \Delta},$ respectively. 
The SOC at time $n$ is written as
\begin{align*}
s(n) = s(0) + s_{\rm CC}(n) + w_\eta(n)
\end{align*}
where
\begin{align}
w_\eta(n)  = & \rho_{\eta_\rc} s_{\rm CCc}(n)  + \rho_{\eta_\rd} s_{\rm CCd}(n)
\end{align}
where is the error in the computed SOC due to the uncertainty in the charging/discharging efficienc.
Similar to $w_{\rm C}$,  $w_\eta(n)$ does not accumulate with time, rather it accumulates with the accumulated Coulombs. 
}

\item {\em Timing oscillator error:}
Considering an error of $\rho_\Delta$ (ratio of clocked time vs. true time) in the timing oscillator, the SOC at time $n$ is derived as 
\begin{align*}
s(n) = s(0) + s_{\rm CC}(n) + w_\Delta(n)
\end{align*}
{
where the SOC error $w_\Delta(n)$ is a deterministic value given by
\begin{align*}
w_\Delta(n) = \rho_\Delta s_{\rm CC}(n)
\end{align*}
}

Similar to the error due to capacity uncertainty, $w_\Delta(n)$ is not  accumulative with time and it is proportional to the accumulated SOC. 
Further, it is shown that practical value of $\eta$ is very small number. 
For example, a timing oscillator that is slower (or faster) by 3 minutes in a month has $\eta = 69 \times 10^{-6}.$ 
Hence, the contribution of timing oscillator error can be considered to be negligible in the computed SOC. 

\end{enumerate}

In summary, the resulting four types of error can be grouped into two categories:
{\em time-accumulative} and {\em SOC-proportional}. 
The SOC errors due to current measurement error and integration approximation fall under the category of time accumulative errors. 
The SOC errors due to the uncertainty in battery capacity and timing oscillator error fall under the category of SOC-proportional errors. 
Next, we briefly discuss the nature of these errors and possible ways to mitigate them.  

\subsection*{Mitigating Time-Accumulative Errors}
{
It must be stressed that the best way to mitigate Coulomb counting errors is to employ a state-space filter, such as the Kalman filter, with correctly derived model parameters --- as briefly discussed in Section \ref{sec:combined}. 
However, practical battery management systems are implemented through complex state diagrams \cite{avvari2015experimental} where at some stages Coulomb counting is the best way to compute the SOC. 
Some strategies discussed below can be useful when the SOC is computed based on Coulomb counting only.
}

The following strategies can be looked at to reduce time-accumulative errors. 
\begin{itemize}
\item {\em Over sampling.}
It can be noted that both $\sigma_{\rm s,i}(n)$ and $\sigma_{\rm s,I}(n)$, in \eqref{eq:sig_s_i_concl} and \eqref{eq:sig_s_I_concl}, respectively, are proportional to $\Delta \sqrt{n}$ where $\Delta$ and $n$ are related by
\begin{align}
n = \frac{T}{\Delta} 
\end{align}
where $T$ is the total time duration. Now, both $\sigma_{\rm s,i}(n)$ and $\sigma_{\rm s,I}(n)$ can be written as
\begin{align}
\sigma_{\rm s,i}(n) &=   \frac{ \rho_{\rm i} }{36 } \sqrt{\Delta T} \quad \%   \\
\sigma_{\rm s,I}(n) &=   \frac{ \kappa \rho_{\rm I} }{36 } \sqrt{\Delta T} \quad \%  
\end{align}
Now, one must realize that the integration error coefficient $\rho_{\rm I}$ reduces with oversampling, i.e., as $\Delta$ decreases so does $\rho_{\rm I}$.
However, the current measurement noise coefficient is unaffected by sampling time.  
The conclusion is that both $\sigma_{\rm s,i}(n) $ and $\sigma_{\rm s,I}(n) $ reduce with higher sampling rate --- however, $\sigma_{\rm s,I}(n) $ reduces at a higher rate compared to $\sigma_{\rm s,i}(n) $ with oversampling. 

\item {\em Reinitialization.}
Time-accumulative errors increase with time. 
Hence, the accumulation of error can be prevented by re-initializing the SOC intermittently. 
For example, the SOC can be reset by OCV-lookup method \cite{balasingam2015performance,avvari2015experimental} where the measured voltage across the battery terminals is used on the OCV-SOC characterization curve in order to find the OCV --- the OCV lookup can be done only when the battery is at rest. 

\end{itemize}

\subsection*{Mitigating SOC-Proportional Errors}

Here, the SOC error is shown to be a fraction of the accumulated SOC over time. 
Intermittent re-initialization --- within a single charge-discharge cycle --- will help to minimize this error. 
However, in most practical cases, there may not be many opportunities (a rested battery) for frequent reset within a single cycle. 
The knowledge of the uncertainty in battery capacity ${\rm \sigma_{batt}}$ will be very useful in the SOC error management. 
For example, if it is known that ${\rm \sigma_{batt}}$ is significantly high, then the SOC can be computed solely based on the voltage approach.

Finally, it must be emphasized that the focus of this paper is exclusively about the Coulomb counting approach. 
As such, we did not delve into other types of approaches that are shown to be useful in improving the SOC estimates, such as the voltage/current based approaches through the use of nonlinear filters \cite{plett2004extended,plett2006sigma}.  
The results reported in this paper, such as the standard deviation of the Coulomb counting error for various scenarios, will help to improve the voltage/current based SOC estimations as well.

{
\section{Combined Effect and the State-Space Model Derivation}
\label{sec:combined}
So far, the Coulomb counting uncertainty is computed only based on individual sources of errors. 
In this section, we discuss how the combined effect due to all sources of error can be approximated using a naive combination approach.
Exact derivation of the combined effect can be quite lengthy due to the non-linear relationships involved --- this is left for a future work. 
Under the naive combination approach, the SOC at time $n$ is written as
\begin{align}
s(n) =s(0) + s_{\rm CC}(n) + w(n)
\end{align}
where
\begin{align}
w(n) = w_{\rm i}(n) + w_{\rm I}(n) + w_{\rm C}(n) + w_\eta(n) + w_\Delta (n)
\label{eq:soc-err-combined}
\end{align}
Under the above naive assumption, it can be shown that 
\begin{align}
E\{w(n) \}  &= 0 \\
E\{w(n)^2 \} = \sigma_{\rm s} (n)^2 & =   
\frac{\Delta^2 \sigma^2_{\rm i} }{3600^2 {\rm C}_{\rm batt}^2} ( \eta_\rc n_\rc + \eta_\rd n_\rd) \nonumber \\
&+ \frac{\kappa^2 \Delta^2  \sigma^2_{\rm L} }{3600^2 {\rm C}_{\rm batt}^2} {( \eta_\rc n_\rc + \eta_\rd n_\rd )}  \nonumber \\
&+ \rho_{\rm C}^2 s_{\rm CC}(n)^2  \nonumber \\
&+  \sigma_{\eta_c}^2 s_{\rm CCc}(n)^2   + \sigma_{\eta_d}^2 s_{\rm CCd}(n)^2  \nonumber \\
&+ \sigma_\Delta^2 s_{\rm CC}(n)^2 
\label{eq:soc-err-combined-var}
\end{align}
With the combined noise derived above, now we are ready to redefine the state-apace model \eqref{eq:process-eq}-\eqref{eq:meas-eq}.  
}

{
Based on the detailed derived about the Coulomb counting error, the process model \eqref{eq:process-eq} can be written as 
\begin{align}
s(k) &= s(k-1) + \frac{\Delta  z_i(k) }{3600 {\rm C}_{\rm batt}} + n_{\rm s} (k) 
\end{align}
where $n_{\rm s} (k) $ is the process noise that has zero-mean and variance given by \eqref{eq:soc-err-combined-var} when $n$ is set to 1. }

{
Based on the notations introduced in \cite{balasingam2014robust}, the measurement equation in \eqref{eq:meas-eq} can be written in detail as follows
\begin{align}
z_\rv(k) = V_\circ(s(k)) + \ba(k)^T \bb + n_z(k)
\label{eq:meas-model-new}
\end{align}
where $V_\circ(s(k)) $ the open circuit voltage model,  
$\ba(k)^T \bb$ approximates the voltage drop in the relaxation elements of the battery, 
$\bb$ the parameter vector of the relaxation elements, 
and
$n_z(k)$ is the measurement noise. 
}

\section{Numerical Analysis}
\label{sec:results}

\subsection{Effect of Current Measurement Error}

\begin{figure}[htbp]
\begin{center}
\includegraphics[width = .75\columnwidth]{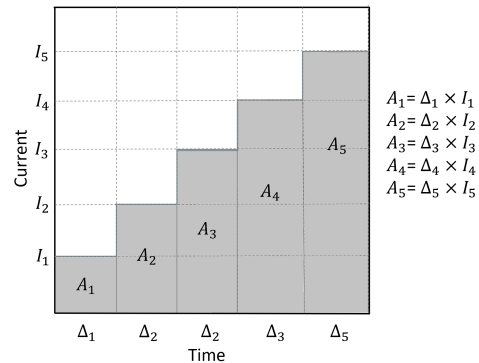}
\caption{
{\bf Generic illustration to computing the true amount of Coulombs.}
Computing true Coulombs is challenging. 
Here, we assume the true current to take the above pattern; under this assumption $ \text{\em Total Coulombs} = A_{1}+A_{2}+A_{3}+A_{4}+A_{5}.$
}
\label{fig:Total_c}
\end{center}
\end{figure}

\begin{figure}[htbp]
\begin{center}
{\includegraphics[width=.75\columnwidth]{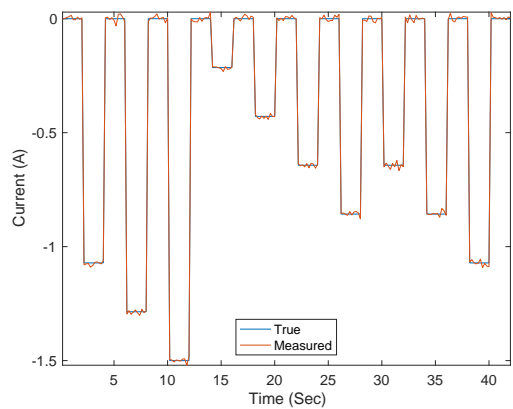}}
\caption{
{\bf Current measurement error.}
True vs. measured current that was simulated by assuming a current measurement error standard deviation of $\sigma_{\rm i} = 10 \,\, {\rm mA}$.
}
\label{fig:z_i}
\end{center}
\end{figure}

\begin{figure}[htbp]
\begin{center}
\subfloat[][Start of load profile]
{\includegraphics[width=.65\columnwidth]{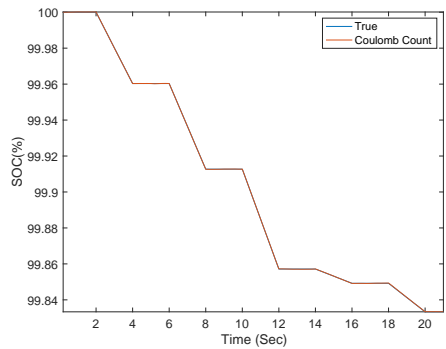}}\\
\subfloat[][End of the load profile in 3.5 hours]
{\includegraphics[width=.65\columnwidth]{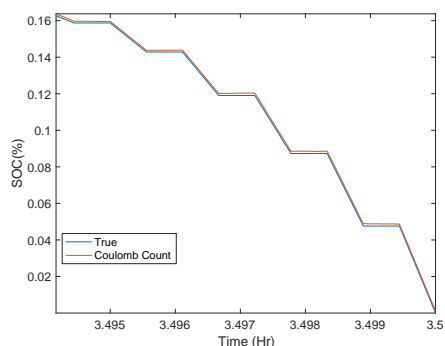}}
\caption{
{\bf Effect of current measurement error in SOC.}
(a) At the start of the experiment, the true SOC and the computed SOC through Coulomb counting are nearly identical. 
(b) Within 3.5 hours, the true SOC and the computed SOC are slightly different. 
{\em Simulation Parameters:} current measurement error s.d. $\sigma_{\rm i} = 10 \,\, {\rm mA}$ and sampling time $\Delta = 200 \,\, {\rm ms.}$
}
\label{fig:SOC:TruevsCC}
\end{center}
\end{figure}

The objective in this section is to validate --- using a Monte-Carlo simulation approach --- the standard deviation of the SOC error due to current measurement error that was derived in \eqref{eq:SOC-Err-SD}. 
For this experiment, errors from all the other possible sources of uncertainties (current integration error, battery capacity uncertainty, timing oscillator error as well as initial SOC error) are assumed to be zero.  
In order to do this, a special current profile, shown in Figure \ref{fig:Total_c}, is created. 
For this profile, the amount of Coulombs can be perfectly computed using geometry.
Once the Coulombs are computed, the true SOC can be computed by making use of the knowledge of the true battery capacity and other noise-free quantities. 
The following procedure details the Monte-Carlo experiment:
\begin{enumerate}
\renewcommand{\theenumi}{\alph{enumi}}
\item 
Generate a perfectly integrable current profile, similar to the one shown in Figure \ref{fig:Total_c}. 
The generated current profile denotes $i(k)$ in \eqref{eq:z_i}.
\item[--] First 40 seconds of the true current profile generated for the experiment is shown Figure \ref{fig:z_i}. 

\item
Compute the true SOC at time $k$, $ s_{\rm true}(k)$, using the Geometric approach illustrated Figure  \ref{fig:Total_c} for the entire duration of the profile, i..e, for $ k=1,\ldots, n$ where $n$ denotes the number of samples in the entire current profile. 

\item
Set $m=1,$ where $m$ denotes the index of the Monte-Carlo run. 

\item 
Generate current measurement noise $n_{\rm i}(k)$ as a zero-mean Gaussian noise with standard deviation $\sigma_{\rm i} = 10 \,\, {\rm mA.}$
Using this, generate the measured current profile $z_{\rm i}(k) = i(k) + n_{\rm i}(k) $. 
\item[--] Figure \ref{fig:z_i} shows the true current profile $i(k)$ along with the measured current profile $z_{\rm i}(k) $ for a duration of 40 seconds. 

\item
Compute the (noisy) SOC, $s_m(k)$ using traditional Coulomb counting equation given in \eqref{eq:CCerror}, i.e., 
$$s_m(k) =s_m(k-1) + \frac{\Delta_k  z_{\rm i}(k) }{3600 {\rm C}_{\rm batt}} $$
where the subscript $m$ denotes the $m^{\rm th}$ Monte-Carlo run. 
\item[--]
Figure \ref{fig:SOC:TruevsCC} shows the true SOC $ s_{\rm true}(k)$ and the computed noisy SOC $s_m(k) $.
The top plot (a) shows the SOC at the start of the current profile and the plot (b) at the bottom shows the SOC towards the end of applying 3.5 hours of load profile. 

\item 
If $m=M$, where $M$ denotes the maximum number of Monte-Carlo runs, go to step g); otherwise,
set $  m \leftarrow m+1 $  and go to step d) 

\item End of simulation (all the data generated during the above steps needs to be stored for analysis). 

\end{enumerate}
After $M=1000$ Monte-Carlo runs, the standard deviation of the SOC error due to current measurement error is computed as 
\begin{eqnarray}
\hat \sigma_{\rm s,i}(k)&=& \sqrt{\frac{1}{M}\sum_{m=1}^M\left(s_{\rm true}(k)- s_m(k)\right)^2}\label{eq:hat_sigma_i}
\label{eq:SOC-SD-MC}
\end{eqnarray}

\begin{figure}[htbp]
\begin{center}
{\includegraphics[width=.75\columnwidth]{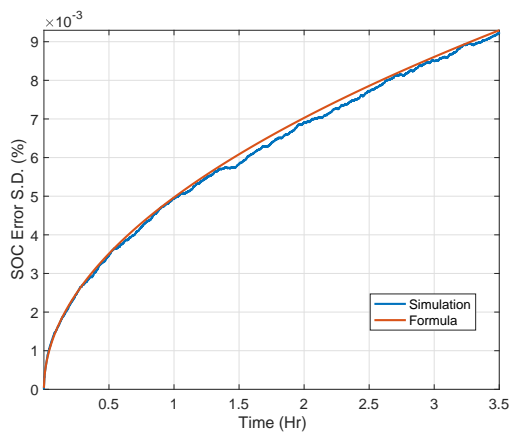}}
\caption{
{\bf Standard deviation of the SOC error due to current measurement error.}
Simulated value is plotted in comparison with the theoretical value derived in \eqref{eq:SOC-Err-SD} shown against time that corresponds to $n$. 
}
\label{fig:SOCerrSD_measErr}
\end{center}
\end{figure}

Figure \ref{fig:SOCerrSD_measErr} shows the standard deviation of the SOC error computed using the theoretical formula \eqref{eq:SOC-Err-SD} and the standard deviation of the SOC error computed using the Monte-Carlo method detailed in \eqref{eq:SOC-SD-MC}. 
As expected, the theoretical derivation matches with the SOC error standard deviation obtained through 1000 Monte-Carlo simulations.

\subsection{Effect of Current Integration Error}

\begin{figure}[htbp]
\begin{center}
\includegraphics[width = .75\columnwidth]{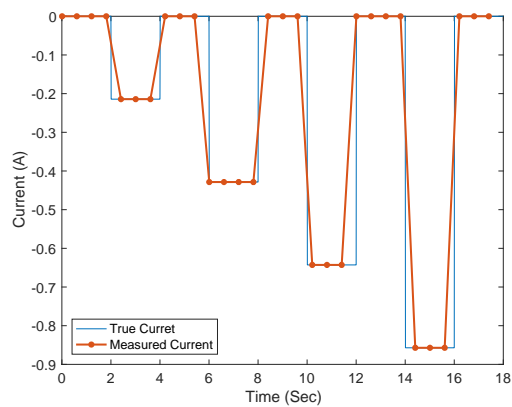}
\caption{
{\bf Perfectly integrable current profile.}
The blue curve shows a perfectly integrable current that is made of rectangular pulses of different amplitude; it can be integrated using the geometric approach detailed in Figure \ref{fig:Total_c}. 
The measured current, shown in red, is a downsampled version of the true current profile -- this emulates the way in which discrete measurement systems measure the voltage/current in BMS. 
}
\label{fig:load_profile_IntErr}
\end{center}
\end{figure}

\begin{figure}[htbp]
\begin{center}
\includegraphics[width = .75\columnwidth]{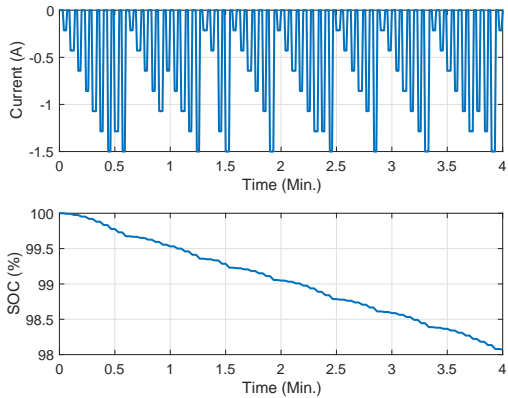}
\caption{
{\bf Current profile and corresponding SOC.}
First four minutes of the true current profile and the corresponding true SOC that is computed using the geometric approach detailed in Figure \ref{fig:Total_c}.
Only 4 minutes of profiles are shown; true profile lasted for 4 hours (see Figure \ref{fig:SOC-Err-SD_IntErr}). 
}
\label{fig:SOC_profile_IntErr}
\end{center}
\end{figure}

The objective in this section is to validate the standard deviation of the SOC error due to integration that we derived in \eqref{eq:sigmainteq}. 
For this experiment, errors from all the other possible sources of uncertainties (current measurement error, battery capacity error, timing oscillator error as well as initial SOC error) are assumed to be zero.  
In order to do this, similar to previous analysis, a special current profile shown in Figure \ref{fig:load_profile_IntErr} is made up of constant current signals of different amplitudes. 
For this profile, the amount of Coulombs can be perfectly computed using geometry similar to the example illustrated in Figure \ref{fig:Total_c}.
Once the Coulombs are computed, the true SOC can be computed by making use of the knowledge of the true battery capacity. 
The following procedure details the Monte-Carlo experiment to validate the standard deviation of the SOC error due current integration error:
\begin{enumerate}
\renewcommand{\theenumi}{\alph{enumi}}
\item 
Generate a perfectly integrable current where the generated current allows one to perfectly compute $\int_{k}^{k+1} i(k) dk$ shown in \eqref{eq:int_error}.
\item[--] 
First 18 seconds of the noiseless current profile $i(k)$ is shown in red Figure \ref{fig:load_profile_IntErr}.
Note that the true current profile is the downsampled version --- this emulates the fact that discretely measured current is always a downsampled version and it will never be the same as the real current (shown in blue).
First four minutes of the current profile along with the true SOC (assuming initial SOC =1) is shown in Figure \ref{fig:SOC_profile_IntErr}. 

\item
Let the true battery capacity to be ${\rm C_{true}} = 1.5 \, {\rm Ah}$.

\item
Assuming the knowledge of the true capacity, compute the true SOC at time $k$, $ s_{\rm true}(k)$, using the geometric approach illustrated Figure \ref{fig:Total_c} for the entire duration of the profile, i..e, for $ k=1,\ldots, n$ where $n$ denotes the number of samples in the entire current profile. 
\item[--] The second plot in Figure \ref{fig:SOC_profile_IntErr} shows the true SOC. 
\item
Set $m=1$ where $m$ denotes the index of the Monte-Carlo run.

\item
Compute the (noisy) SOC $s_m(k)$ using traditional Coulomb counting equation given in \eqref{eq:CCerror}, i.e., 
$$s_m(k) =s_m(k-1) + \frac{\Delta_k  i(k) }{3600 {\rm C}_{\rm batt}} $$
where 
$i(k)$ are the `measured current' indicated by red lines in Figure \ref{fig:load_profile_IntErr},
and the subscript $m$ denotes the $m^{\rm th}$ Monte-Carlo run.

\item 
If $m=M$, where $M$ denotes the maximum number of Monte-Carlo runs, go to step g); otherwise,
set $  m \leftarrow m+1 $  and go to step e) 

\item End of simulation (all the data generated during the above steps needs to be stored for analysis). 

\end{enumerate}
After $M=1000$ Monte-Carlo runs, the standard deviation of the SOC error due to current measurement error is computed as 
\begin{eqnarray}
\hat \sigma_{\rm s,I}(k)&=& \sqrt{\frac{1}{M}\sum_{m=1}^M\left(s_{\rm true}(k)- s_m(k)\right)^2}\label{eq:hat_sigma_i}
\label{eq:SOC-SD-MC-IntError}
\end{eqnarray}

\begin{figure}[htbp]
\begin{center}
\includegraphics[width = .75\columnwidth]{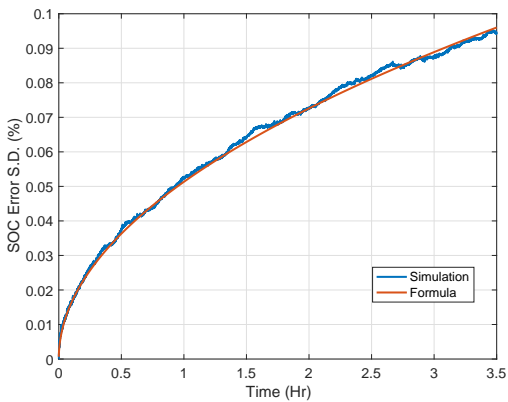}
\caption{
{\bf The Standard deviation of SOC error due to current integration error.}
The red curve is the theoretical value of the s.d. $\sigma_{\rm s,I}(k)$ derived in \eqref{eq:SOC-Err-SD-IntErr}; 
the blue curve shows $\hat \sigma_{\rm s,I}(k)$, the s.d. obtained through Monte-Carlo simulation as shown in \eqref{eq:SOC-SD-MC-IntError}.
The constant $\kappa$ is computed through empirical methods to be $\kappa=.88$. 
It must be noted that $\kappa$ varies for different types of current profiles. 
}
\label{fig:SOC-Err-SD_IntErr}
\end{center}
\end{figure}

Figure \ref{fig:SOC-Err-SD_IntErr} shows the standard deviations of error computed through the theoretical approach, $\sigma_{\rm s,I}(n)$ in \eqref{eq:SOC-Err-SD-IntErr}, and through the Monte-Carlo simulation approach, $\hat \sigma_{\rm s,I}(k)$ \eqref{eq:SOC-SD-MC-IntError}. 
The constant $\kappa$ for the theoretical approach in \eqref{eq:SOC-Err-SD-IntErr} is found to be $\kappa = 0.88$ through empirical means (i.e., different values for $\kappa$ was used until the theoretical curve in red aligned well with the simulation curve in blue). 
It must be noted that  $\kappa$ will be different for different current profiles.

\subsection{Effect of Battery Capacity Uncertainty}

\begin{figure}[htbp]
\begin{center}
\includegraphics[width = .75\columnwidth]{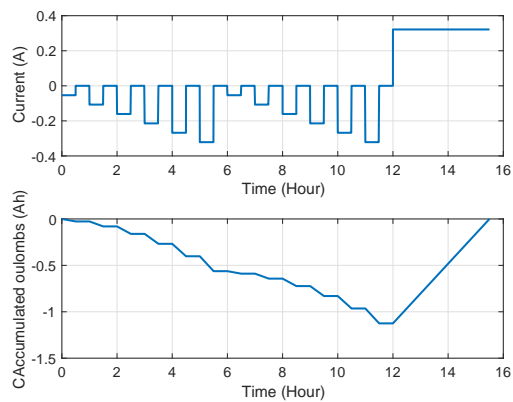}
\caption{
{\bf Simulated current profile and corresponding true SOC.}
This figure is showing the difference between the true SOC and the SOC with battery capacity uncertainty after 100 runs of Monte Carlo.
}
\label{fig:cbatt_er}
\end{center}
\end{figure}

\begin{figure}[htbp]
\begin{center}
\includegraphics[width = .75\columnwidth]{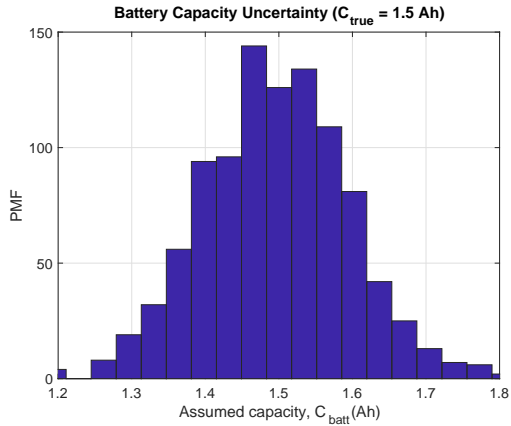}
\caption{
{\bf The histogram of ${\rm C_{batt}}$} generated during 1000 Monte Carlo simulations.
This graph is showing that the battery capacity error that we are using in our Monte Carlo runs is reasonable.
}
\label{fig:cbatt_hist}
\end{center}
\end{figure}

\begin{figure}[htbp]
\begin{center}
\includegraphics[width = .75\columnwidth]{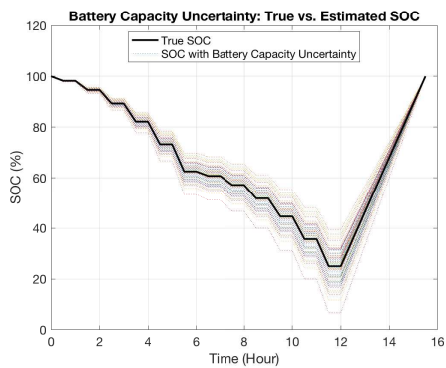}
\caption{
{\bf SOC error due to battery capacity uncertainty.}
This figure is showing the difference between the true SOC and the SOC with battery capacity uncertainty for different simulation.
The true SOC is computed using the true battery capacity of ${\rm C_{true} = 1.5 \,\, Ah}$;
Each Monte Carlo run assumes a different battery ${\rm C_{batt}}$ that is distributed $N({\rm C_{true}}, \sigma^2_{\rm batt})$.
Figure \ref{fig:cbatt_hist} all the ${\rm C_{batt}}$ during different runs. 
}
\label{fig:SOC_err_cbatt}
\end{center}
\end{figure}

The objective in this section is to validate the standard deviation of the SOC error due to battery capacity uncertainty that we derived in \eqref{eq:STD-Cbatt-error} using Monte-Carlo simulation approach. 
For this experiment, errors from all the other possible sources of uncertainties (current measurement error, current integration error, timing oscillator error as well as initial SOC error) are assumed to be zero.  
In order to do this, similar to previous analysis, a special current profile that is shown in Figure \ref{fig:cbatt_er} is created. 
The current profile in Figure \ref{fig:cbatt_er} is made of low frequency (constant current) signals of different amplitudes. 
For this profile, the amount of Coulombs can be perfectly computed using geometry similar to the example illustrated in Figure \ref{fig:Total_c}.
Once the Coulombs are computed, the true SOC can be computed by making use of the knowledge of the true battery capacity. 
The following procedure is followed to perform the Monte-Carlo experiment to validate the standard deviation of the SOC error due to uncertainty in battery capacity:
\begin{enumerate}
\renewcommand{\theenumi}{\alph{enumi}}
\item 
Generate a perfectly integrable current where the generated current profile denotes $i(k)$ in \eqref{eq:z_i}.
\item[--] The entire true current profile generated for the experiment is shown at the top plot Figure \ref{fig:cbatt_er}. 

\item
Let the true battery capacity to be ${\rm C_{true}} = 1.5 \, {\rm Ah}$.

\item
Assuming the knowledge of the true capacity, compute the true SOC at time $k$, $ s_{\rm true}(k)$, using the geometric approach illustrated Figure \ref{fig:Total_c} for the entire duration of the profile, i..e, for $ k=1,\ldots, n$ where $n$ denotes the number of samples in the entire current profile. 
\item[--] The second plot in Figure \ref{fig:cbatt_er} shows the accumulated Coulombs $s_{\rm CC}(n)$. From this, the true SOC can be computed as $s_{\rm true}(n) = s(0) + s_{\rm CC}(n).$

\item
Set $m=1$ where $m$ denotes the index of the Monte-Carlo run.

\item 
Assuming capacity estimation error s.d. of $\sigma_{\rm batt} = 0.1 \, {\rm Ah}$ use the capacity uncertainty model of \eqref{eq:cbatt-ctrue} to compute the estimate battery capacity ${\rm C_{batt}} = {\rm C_{true}} + {\rm C_{\Delta}}$ where is a zero-mean random number with standard deviation $\sigma_{\rm batt}.$
\item[--] Figure \ref{fig:cbatt_hist} shows all the ${\rm C_{batt}}$ values generated for $m=1,\ldots, M$ in the form of a histogram. 

\item
Compute the (noisy) SOC $s_m(k)$ using traditional Coulomb counting equation given in \eqref{eq:CCerror}, i.e., 
$$s_m(k) =s_m(k-1) + \frac{\Delta_k  i(k) }{3600 {\rm C}_{\rm batt}} $$
where the subscript $m$ denotes the $m^{\rm th}$ Monte-Carlo run. 
\item[--]
Figure \ref{fig:SOC_err_cbatt} shows the true SOC $ s_{\rm true}(k)$ and the computed noisy SOC $s_m(k) $ for different Monte-Carlo runs.

\item 
If $m=M$, where $M$ denotes the maximum number of Monte-Carlo runs, go to step h); otherwise,
set $  m \leftarrow m+1 $  and go to step e) 

\item End of simulation (all the data generated during the above steps needs to be stored for analysis). 

\end{enumerate}
After $M=1000$ Monte-Carlo runs, the standard deviation of the SOC error due to current measurement error is computed as 
\begin{eqnarray}
\hat \sigma_{\rm s,C}(k)&=& \sqrt{\frac{1}{M}\sum_{m=1}^M\left(s_{\rm true}(k)- s_m(k)\right)^2}\label{eq:hat_sigma_i}
\label{eq:SOC-SD-MC-Cbatt}
\end{eqnarray}

\begin{figure}[htbp]
\begin{center}
\includegraphics[width = .75\columnwidth]{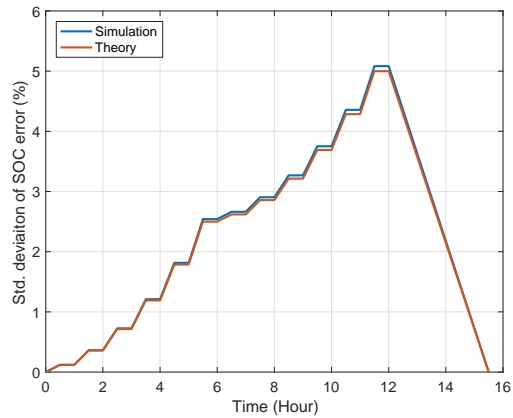}
\caption{
{\bf The Standard deviation of SOC error due to battery capacity uncertainty.}
The red curve is the theoretical value of the s.d. $\sigma_{\rm s,C}(k)$ derived in \eqref{eq:STD-Cbatt-error}; the blue curve shows $\hat \sigma_{\rm s,C}(k)$, the s.d. obtained through Monte-Carlo simulation as shown in \eqref{eq:SOC-SD-MC-Cbatt}.
Due to the approximation made in \eqref{eq:TaylorApprox}, the theoretical and simulated values slightly differ. 
}
\label{fig:soc_err_sd_cbatt}
\end{center}
\end{figure}

Figure \ref{fig:soc_err_sd_cbatt} shows the SOC error standard deviation obtained through the theoretical equation \eqref{eq:STD-Cbatt-error} as well as the Monte-Carlo simulation approach summarized through \eqref{eq:SOC-SD-MC-Cbatt}. 
It can be noticed that the theoretical value and the simulated values slightly differ --- this can be attributed to the approximation made in \eqref{eq:TaylorApprox} in order to derive the theoretical value. 

%{
%\subsection{Effect of Charging Efficiency Error}
%{I will include this - Bala.}
%%
%%
%\subsection{Effect of Timing Oscillator Error}
%{I will include this - Bala.}
%%
%%
%%
%\subsection{Demonstration of the new State-Space Model}
%{I will include this - Bala.}
%}
%

\section{Conclusions and Discussions}
\label{sec:concl}
In this paper, we developed an in-depth mathematical analysis of Coulomb counting method for state of charge estimation in rechargeable batteries. 
Particularly, we derived the exact statistical values of the state of charge error as a result of 
(i) current measurement error,
(ii) current integration error,
(iii) battery capacity uncertainty,
and 
(iv) timing oscillator error. 
It was shown that the state of charge error due to current measurement error  and  current integration error grow with time whereas the 
state of charge error due to battery capacity uncertainty and timing oscillator error are proportional to the accumulated state of charge that ranges between 0 and 1. 
The models presented in this paper will be useful to improve the overall state of charge estimation in majority of the existing approaches. 

% talk about combined effects - that needs to be defined based on the assumptions

\section*{Acknowledgments}
\label{sec:acknowledge}
B. Balasingam acknowledges the support of the Natural Sciences and Engineering Research Council of Canada (NSERC) for financial support under the Discovery Grants (DG) program [funding reference number RGPIN-2018-04557].
B. Balasingam acknowledges the help of Mostafa Ahmed and Arif Raihan for their help searching through relevant literature for Section I of this manuscript. 
Research of K. Pattipati was supported in part by the U.S. Office of Naval Research and US Naval Research Laboratory under Grants
\#N00014-18-1-1238, \#N00173-16-1-G905, \#HPCM034125HQU and by a Space Technology Research Institutes grant (\#80NSSC19K1076) from NASA's Space Technology Research Grants Program.

%
%\bibliographystyle{ieeetr}
%\bibliography{BFGref1,BFGref2}

\end{document}